\documentclass[]{emulateapj}

\def\rmit#1{{\it #1}}              
\def\specchar#1{{\sc #1}}
\def\FeI{\mbox{Fe\,\specchar{i}}}

\def\SiI{\mbox{Si\,\specchar{i}}}

\def\CaII{\mbox{Ca\,\specchar{ii}}}

\def\ie{\rmit{i.e.}}
\def\eg{\rmit{e.g.}}

\usepackage{natbib,graphicx}

\usepackage{color}

\shorttitle{Polarimetric data from a sunspot simulation}

\shortauthors{Felipe et al.}

\begin{document}

\title{Synthetic observations of wave propagation in a sunspot umbra}

\author{T. Felipe\altaffilmark{1}, H. Socas-Navarro\altaffilmark{2,3}, E. Khomenko\altaffilmark{2,3}}
\email{tobias@nwra.com}

\altaffiltext{1}{NorthWest Research Associates, Colorado Research Associates, Boulder, CO 80301, USA}
\altaffiltext{2}{Instituto de Astrof\'{\i}sica de Canarias, 38205,
C/ V\'{\i}a L{\'a}ctea, s/n, La Laguna, Tenerife, Spain}
\altaffiltext{3}{Departamento de Astrof\'{\i}sica, Universidad de La Laguna, 38205, La Laguna, Tenerife, Spain}

\begin{abstract}
Spectropolarimetric temporal series from \FeI\ $\lambda$ 6301.5 \AA\ and \CaII\ infrared triplet lines are obtained by applying the Stokes synthesis code NICOLE to a numerical simulation of wave propagation in a sunspot umbra from MANCHA code. The analysis of the phase difference between Doppler velocity and intensity core oscillations of the \FeI\ $\lambda$ 6301.5 \AA\ line reveals that variations in the intensity are produced by opacity fluctuations rather than intrinsic temperature oscillations, except for frequencies between 5 and 6.5 mHz. On the other hand, the photospheric magnetic field retrieved from the weak field approximation provides the intrinsic magnetic field oscillations associated to wave propagation. Our results suggest that this is due to the low magnetic field gradient of our sunspot model. The Stokes parameters of the chromospheric \CaII\ infrared triplet lines show striking variations as shock waves travel through the formation height of the lines, including emission self-reversals in the line core and highly abnormal Stokes V profiles. Magnetic field oscillations inferred from the \CaII\ infrared lines using the weak field approximation appear to be related with the magnetic field strength variation between the photosphere and the chromosphere.

\end{abstract}



\section{Introduction}

Magnetic fields play a key role in the structure and dynamics of the solar atmosphere. Sunspots and active regions are the most prominent manifestations of the magnetic activity in the solar surface, but even the quiet Sun contains ubiquitous magnetic fields \citep{Trujillo-Bueno+etal2004}. The analysis of spectropolarimetric observations is the most common approach to study the properties of the magnetized atmosphere \citep[see][for a review]{Solanki1993, Stenflo2013}. A detailed interpretation of the Stokes profiles is a strong requirement to determine the three-dimensional structure of the solar magnetic field and unveil the coupling between different layers and the phenomena associated with it.

Among all those phenomena, the heating of the outer atmosphere is one of the most remarkable and long-standing unanswered problems in solar physics. Several mechanisms have been proposed to explain the coronal heating \citep[see][for a review]{Klimchuk2006}. The most likely candidates are wave heating \citep{Alfven1947, Biermann1948, Schwarzschild1948} and magnetic reconnection \citep{Parker1983, Heyvaerts+Priest1983}. Many efforts of the solar physics research community aim to find observables associated with those mechanisms. The study of coronal heating is a main target of current infrastructures like Hinode satellite \citep{Kosugi+etal2007} and future projects like the Advanced Technology Solar Telescope \citep{Keil+etal2003} or Solar-C. In order to fully exploit the high quality data from state-of-the-art observations, it is necessary to understand what processes are we observing and perform an accurate interpretation of the measurements.

Velocity and intensity oscillations are easily observed using spectral lines and, thus, have been studied in depth. However, retrieving magnetic field oscillations is an observational challenge and need the use of sophisticated analysis methods. Using full Stokes inversions of the \FeI\ $\lambda$ 6301.5 \AA\ and \FeI\ $\lambda$ 6302.5 \AA\ lines, \citet{Lites+etal1998} found an upper limit of 4 G for the amplitude of 5 minute oscillations in magnetic field strength, and considered them to be an instrumental artifact rather than intrinsic solar oscillations. \citet{Ruedi+etal1998} reported highly localized magnetic oscillations in different umbral regions with variations around 6 G. \citet{BellotRubio+etal2000} detected amplitudes around 7-11 G, but according to the phase lag between the oscillations in the line-of-sight velocity and the magnetic field they suggested that the measured magnetic field oscillations are produced by opacity effects, which shift the height were spectral lines are sensitive to magnetic field. \citet{Khomenko+etal2003} compared the observations from \citet{BellotRubio+etal2000} with an analytical solution of the MHD equations using a model which accounts for gravity, inclination of the magnetic field, and effects of nonadiabaticity and concluded that the observed oscillations are partly due to real variations of the magnetic field produced by magnetoacoustic wave modes. \citet{Fujimura+Tsuneta2009} reported fluctuations of the magnetic field with amplitude of 4-17 G in the photosphere of magnetic flux tubes. Based on the phase relation between the magnetic field strength and intensity oscillations, they discarded the magnetic fluctuations to be caused by opacity effects, and they suggested that they are due to sausage and kink waves.

In this paper, we attempt to evaluate the Stokes profiles expected from the propagation of magnetoacoustic waves between the photosphere and chromosphere of sunspots, and address what information can be extracted from them. With this aim, we have synthesized several spectral lines commonly used in observations in a numerical simulation of wave propagation in a sunspot umbra. The rest of the paper is organized as follows: Section \ref{sect:simulation} describes the numerical simulation; Section \ref{sect:synthesis} is devoted to the analysis of the synthetic spectra, including the description of the synthesis code (Section \ref{sect:nicole}), the analysis of the \FeI\ $\lambda$ 6301.5 \AA\ line (Section \ref{sect:Fe}), and the analysis of the \CaII\ infrared (IR) triplet (Section \ref{sect:Ca}); and finally, we summarize the results in Section \ref{sect:conclusions}.

\section{3D MHD numerical simulation}
\label{sect:simulation}
We use a MHD simulation of wave propagation in a sunspot umbra computed using the code MANCHA \citep{Khomenko+Collados2006, Felipe+etal2010a}. The code solves the three-dimensional (3D) MHD equations for perturbations, which are obtained by removing the equilibrium state from the equations. The computational domain is discretized using a 3D Cartesian grid with constant space step in each dimension. The spatial derivatives are discretized using five grid points in a fourth-order centered differences scheme, and the solution is advanced in time by an explicit fourth-order Runge-Kutta. Following \citet{Vogler+etal2005}, the physical diffusive terms in the momentum, induction, and energy equations are replaced by artificial equivalents in order to damp high-frequency numerical noise on sub-grid scales. Perfect Matched Layers \citep{Berenger1996} are used at all the boundaries in order to absorb waves without producing reflections. Radiative transfer was implemented following Newton's cooling law, while thermal conduction was neglected because its time scale is several orders of magnitude higher than that of radiation.

The simulation analyzed in this work was designed to reproduce the observations presented in \citet{Felipe+etal2010b}. A magnetohydrostatic sunspot model based of the properties of the observed sunspot was constructed following the method described in \citet{Khomenko+Collados2008}. This method generates a thick sunspot atmosphere in magnetostatic equilibrium with distributed currents, where the magnetic field and the thermodynamic magnitudes change smoothly from the axis of the sunspot to the quiet Sun atmosphere at large radial distances. For this simulation, we only used the central part of the model, corresponding to the umbra. Photospheric oscillations were driven by introducing the fluctuations measured with the photospheric \SiI\ $\lambda$ 10827 \AA\ line at its corresponding formation height. In order to generate photospheric oscillations as close as possible to those measured with the \SiI\ $\lambda$ 10827 \AA\ line, we chose to introduce a source function in the momentum equation. Note that in this simulation we use the internal energy equation (instead of imposing the conservation of the total energy) and, thus, it is not necessary to add a source term to the energy equation since the energy of the driver should not be employed in heating the plasma. The characteristics of this driving force, the details of its implementation, and the justification the convenience of this approach are described in \citet{Felipe+etal2011}. 

Pure fast and slow magneto-acoustic modes only exist in homogeneous magnetic fields, unlike the sunspot structure described in the paper. In such an inhomogeneous atmosphere the slow, fast, and Alfv\'en modes are coupled, especially near the deepest layers of our model where the $\beta$ parameter of the plasma is closer to unity. Although in realistic atmospheres there is no clear mathematical division between pure wave modes, we borrowed these definitions from theory because their simplicity provides a good approximation for describing the simulated waves. Since the driver introduces a vertical force in a region dominated by a vertical magnetic field ($\beta$ below 1), it mainly generates a longitudinal wave whose behavior resembles that of a slow mode. In the following, we refer to this idealized picture when we use the terms ``slow and fast magneto-acoustic modes''.  

The simulation domain covers $14.8\times 8.4$ Mm in the horizontal directions, with a spatial step of $\Delta x=\Delta y=0.1$ Mm, while in the vertical direction it spans from $z=-0.6$ Mm to $z=1$ Mm with $\Delta z=0.025$ Mm, resulting in a grid size of $148\times 84\times 64$. The layer $z=0$ is located at the height where optical depth at 5000 \AA\ $\tau_{5000}=1$ in the non-magnetic atmosphere. Note that due to the 0.35 Mm of Wilson depression of the model, the top boundary is located at $1.35$ Mm above the photospheric height in the umbra. The duration of the simulation is 72 minutes.  The details of the calculations and the comparison of the numerical data with the real observations from \citet{Felipe+etal2010b} can be found in \citet{Felipe+etal2011}. 

It is well known that the solar photosphere and chromosphere are regions
with a very low degree of atomic ionization, reaching the minimum in the
photosphere of about $10^{-4}$. In these regions the plasma is strongly
collisionally coupled, in a way that the collision frequency is orders of
magnitude larger than the  gyro-frequency of ions and electrons. As it
becomes clear from the recent studies \citep{Vranjes+Poedts2008, Soler+etal2013, Khomenko+Collados2012}, non-ideal plasma effects derived
from the presence of a large amount of neutrals must be taken into account
for the description of many phenomena related to the magnetic field and
energy release, even in the photosphere and chromosphere. For such strongly
collisionally coupled plasma, a single-fluid description together with the
generalized Ohm's law is usually sufficient to apply. An order of magnitude
analysis based on the evaluation of the Ambipolar and Hall terms (most
important non-ideal terms in the generalized Ohm's law) shows that the
spatial and temporal scales at which non-ideal effects become important are
of the order of 10-100 m and 0.1 - 0.001 sec at maximum. Since in our
simulations we deal with scales of the order of a few km (grid resolution)
and few tens or hundreds of sec (typical wave period), we expect that the
effect of the non-ideal terms is not large. Nevertheless, in the
chromosphere, where strong shocks are formed in our simulations, the
non-ideal effects may become important at the shock fronts and would lead
to additional energy dissipation and release. Such effects are beyond
the scope of the current paper focused primarily at the observational
diagnostics, since typical observational resolution is significantly larger
than that of numerical simulations. But they are definitely worth
considered in a specially dedicated study.

\begin{figure*}[!ht] 
 \centering
 \includegraphics[width=14cm]{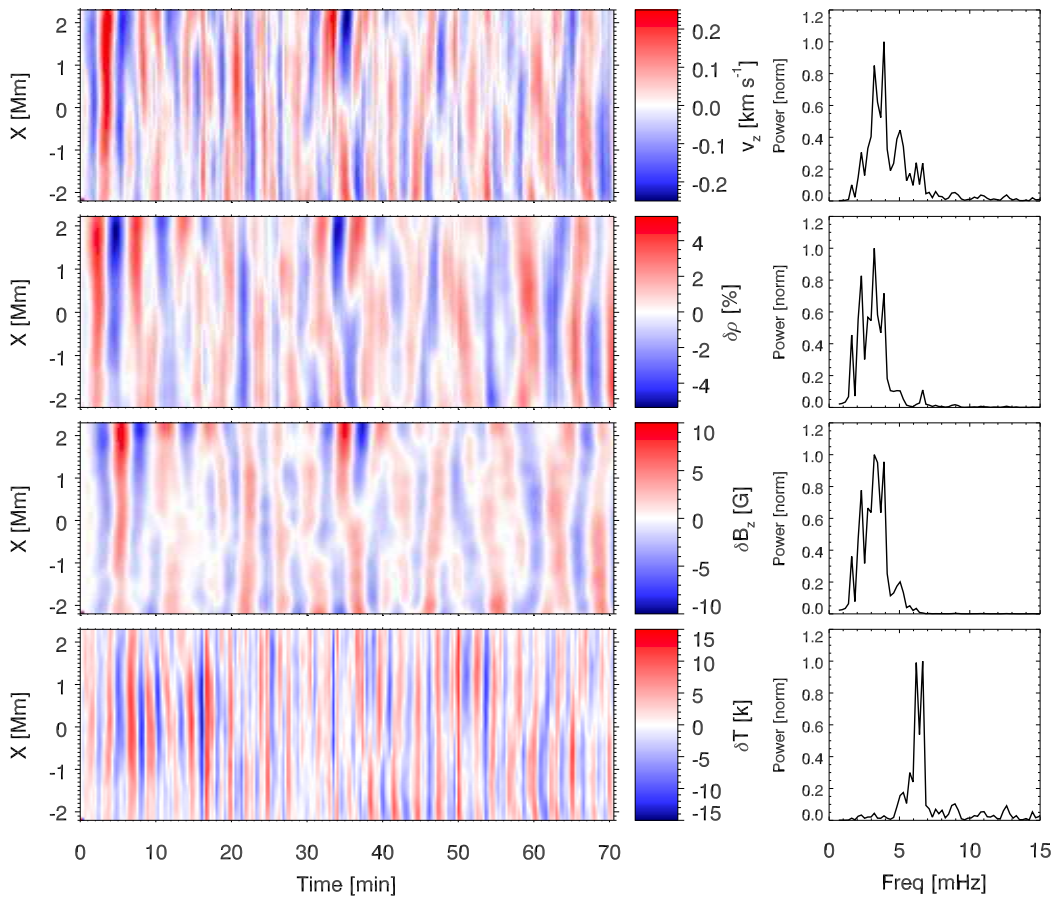}
  \caption{Overview of the simulation at the photosphere, from top to bottom: velocity, density, vertical magnetic field, and temperature perturbations. Left panels: spatial-velocity maps. Right panels: normalized power spectra. }
  \label{fig:maps_photosphere}
\end{figure*}

Figure \ref{fig:maps_photosphere} shows the temporal evolution of the oscillations in velocity, density, vertical magnetic field, and temperature in the inner part of the umbra and their power spectra at $z=-0.10$ Mm and $y=0$ Mm. Note that this simulation was designed to reproduce an observed temporal series. In these observations we only know the wave field along the slit of the spectrograph (placed along the $x$ direction in the simulation) for a single $y$ position at the center of the sunspot. This way, the driver only covers a chosen thickness in the $y$ direction and it is smoothly modulated to zero after a few grid points. However, the simulations are computed in full 3D because we want the energy to be distributed in a three-dimensional geometry. In the following, all figures and discussions are refereed to the plane $y=0$. Fluctuations in density, magnetic field, and temperature correspond to departures from the MHS equilibrium state, obtained as the difference between their total value (with spatial and temporal variations) and the background. The background has no velocity flows, so the total velocity is the perturbed value. Positive values are plotted as red colors and indicate an increase in the magnitude. In the case of the velocity, they correspond to downflows.

The power of velocity, density, and magnetic field fluctuations is concentrated in the 5 minutes band. Their power spectra show several power peaks at the same frequencies for the three magnitudes. The relative strength of the peak differs between them, but the cause of this is not clear. The highest power peak of the velocity is located at 3.90 mHz, while for the density and magnetic field it is at 2.21 mHz. In general, the distribution of the power of density oscillations is more prominent at lower frequencies, and the velocity power extends to higher frequencies, including some peaks above the cut-off frequency. The cut-off frequency is computed as:

\begin{equation}
w_{\rm c}=\frac{c_{\rm S}}{2H_{\rm \rho}}\sqrt{1-2\frac{dH_{\rm \rho}}{dz}},
\label{eq:Fac}
\end{equation}

\noindent where $c_{\rm S}$ is the sound speed and $H_{\rm \rho}(z)$ is the density scale height. It changes with height. Its maximum value of 6.33 mHz is found at $z=0.175$ Mm. Higher frequency waves propagate towards higher layers and dominate at the chromosphere as a result of their large amplitude increase in comparison with the low frequency evanescent waves \citep{Centeno+etal2006, Felipe+etal2010b}. On the other hand, temperature fluctuations show no power in the 5 minute band. This is due to the relation between the radiative relaxation time and the period of the waves. In the case of long period waves, at the photosphere the time scale of the radiation is significantly shorter than their period and temperature oscillations do not capture the fluctuations of the waves. On the other hand, waves in the 3 minute band have shorter periods, closer to the radiative relaxation time, and show some oscillatory power.

Table \ref{tab:simulations} shows a summary of the oscillations of several variables obtained at the photosphere ($z=-0.10$ Mm, upper index $ph$) and the chromosphere ($z=0.80$ Mm, upper index $ch$), including their rms amplitude, the rms amplitude of the lower and higher frequency oscillations, and the maximum amplitude. The magnitude of the density oscillations is given as a percentage with respect to the equilibrium value of the density at the corresponding height. At the photosphere the lower frequency oscillations (which includes the 5 minutes band) show higher rms amplitudes except for the temperature, whose fluctuations almost vanish. The absence of oscillatory behavior in the temperature at the photosphere is consistent with observational works \citep{Lites+etal1998,BellotRubio+etal2000}. Note that this simulation reproduces an observed time series \citep{Felipe+etal2011}, so the tabulated values of velocity, density, magnetic field, and temperature are expected to show a quantitative agreement with actual observations of the propagation of magnetoacoustic waves in sunspot umbrae. 

\begin{figure*}[!ht] 
 \centering
 \includegraphics[width=14cm]{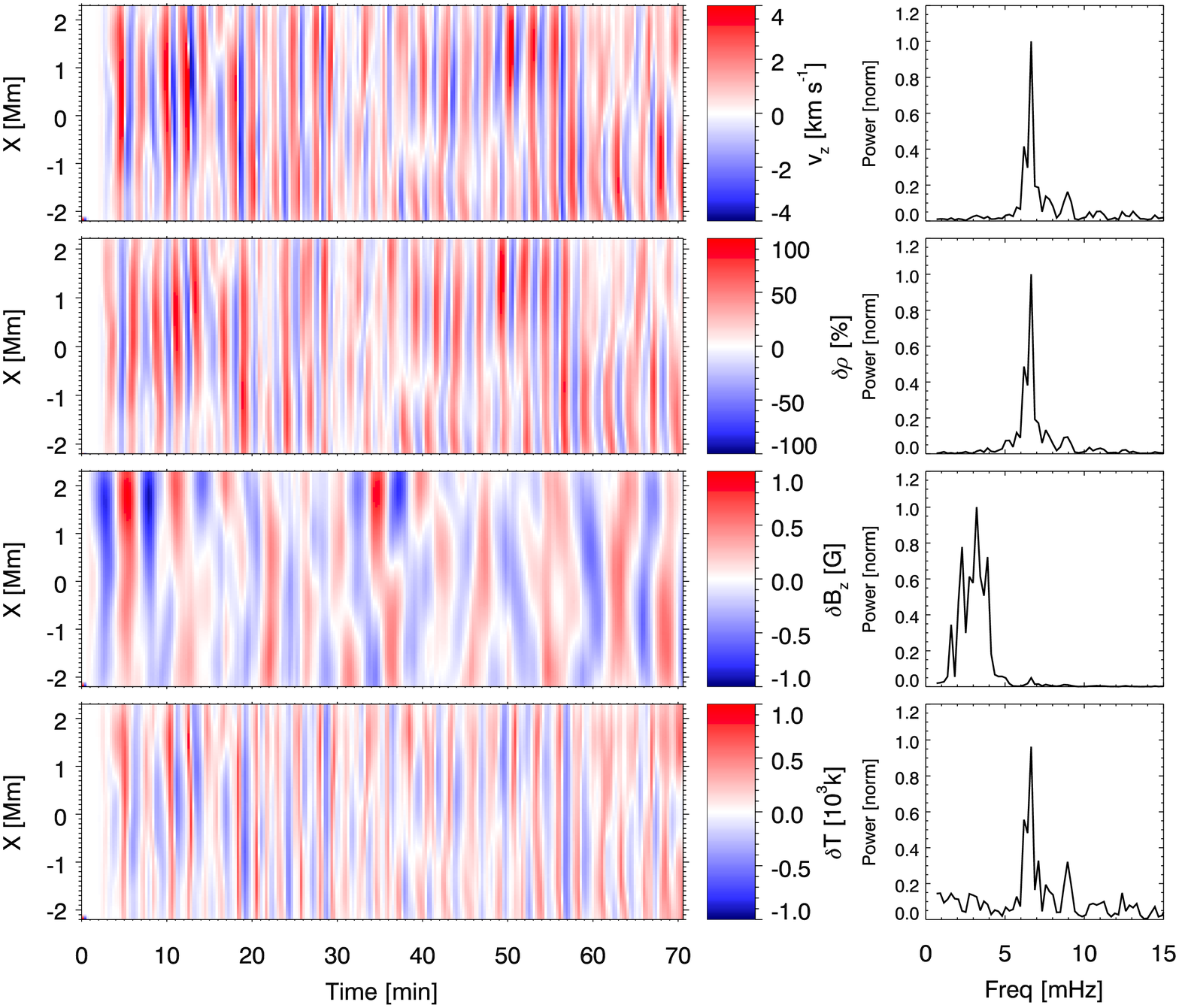}
  \caption{Overview of the simulation at the chromosphere, from top to bottom: velocity, density, vertical magnetic field, and temperature perturbations. Left panels: spatial-velocity maps. Right panels: normalized power spectra. }
  \label{fig:maps_chromosphere}
\end{figure*}

Figure \ref{fig:maps_chromosphere} shows spatio-temporal maps and power spectra of the same variables at the chromosphere, at $z=0.8$ Mm. The power of velocity, density, and temperature fluctuations is located at the 3 minutes band, and the peak of all of them is at 6.60 mHz. The frequency of this peak corresponds to the frequency above the cut-off with more oscillatory power at the photosphere, where waves are driven. As the propagating high frequency waves reach higher layers the amplitude of their velocity oscillations increases due to the density falloff. As can be seen in the top panel of Figure \ref{fig:maps_chromosphere}, they present peak-to peak amplitudes up to 8 km s$^{-1}$ and develop into shocks. The shocks are followed by an increase in the density. After the strongest shocks, the density can be as high as more than twice the equilibrium value (density perturbation takes values up to 103.6\%, Table \ref{tab:simulations}), and then it is reduced to around half the equilibrium value. As shown in Table \ref{tab:simulations}, the rms amplitude of the higher frequencies of those variables in the chromosphere is strikingly higher than that for the evanescent lower frequency waves. However, magnetic field oscillations show a completely different behavior. At the chromosphere, their power is concentrated in the 5 minutes frequency band and their amplitude has been reduced an order of magnitude. Magnetic field fluctuations at the low $\beta$ chromospheric atmosphere are essentially produced by the fast magnetoacoustic mode, which is affected in a different way by the cut-off frequency. Low frequency fast magnetoacoustic waves can propagate to higher layers and, despite their low amplitude due to the small excitation of the fast mode by the driver, magnetic oscillations in the 5 minutes frequency band dominate at the chromosphere.

\begin{table}[t]
\begin{center}
\caption[]{\label{tab:simulations}
          { Summary of the simulation}}
\begin{tabular*}{8cm}{@{\extracolsep{\fill}}ccccc}

\hline Variable              & rms    & rms             & rms           & max  \\
                             &        & [$\nu<4.5$ mHz] & [$\nu>5$ mHz] &  ampl \\

 \hline
  $v_z^{ph}$ (km s$^{-1}$)   &  0.075 &      0.057         &    0.036             &  0.3         \\
  $\delta \rho^{ph}$ (\%)   &  1.5   &        1.3         &    0.4            &  5.4       \\
  $B_z^{ph}$ (G)            &  2.1   &      2.0           &    0.5             &  9.9         \\
  $\delta T^{ph}$ ($k$)     &  4.3   &      0.84          &    3.5            &   16        \\
  $v_z^{ch}$ (km s$^{-1}$)   &  1.4   &       0.3          &    1.2             &  4.9         \\
  $\delta \rho^{ch}$ (\%)  &  30.1  &        5.5         &    27.4            &  103.6       \\
  $B_z^{ch}$ (G)            &  0.27   &      0.25           &   0.05             &  0.98         \\
  $\delta T^{ch}$ ($k$)    &  281.9 &      100.3         &    190.4            &   1095        \\

\hline

\end{tabular*}
\end{center}
\end{table}

In order to evaluate the wave energy introduced in the simulation, we have computed the wave energy fluxes following \citet{Bray+Loughhead1974}. The averaged acoustic energy flux is given by:

\begin{equation}
{\bf F_{ac}}=\langle p_1{\bf v}\rangle,
\label{eq:Fac}
\end{equation}

\noindent while the averaged magnetic energy flux is calculated from the expression: 

\begin{equation}
{\bf F_{mag}}=\langle {\bf B_1}\times({\bf v}\times {\bf B_0})/\mu_0\rangle.
\label{eq:Fmag}
\end{equation}

\noindent where $p_1$, ${\bf v}$, and ${\bf B_1}$ are the perturbations in pressure, velocity, and magnetic field, respectively, ${\bf B_0}$ is the background magnetic field, and $\mu_0$ is the magnetic permeability. The brackets indicate the averaging of the fluxes for all the spatial positions inside the umbra and all the time steps after the first wavefront reaches the chromosphere. The variation of the averaged vertical acoustic flux with height can be found in Figure 16 from \citet{Felipe+etal2011}. At the location of the driver ($z=-0.1$ Mm) it vanishes because it represents the average of the downward and upward fluxes, and they cancel out according to the method used for introducing the driver. At $z=0$, the averaged vertical acoustic energy flux is $1.3\times 10^6$ erg cm$^{-2}$s$^{-1}$ directed towards higher layers. The average magnetic flux is around two orders of magnitude lower, and it is more significant in the horizontal direction. Most of the energy of the driver goes to the acoustic wave flux and, together with the fact that at this height the atmospheric model is dominated by magnetic pressure, supports our previous statement that we are exciting slow-like magneto-acoustic waves. The horizontal magnetic wave flux shows higher values out of the axis of the sunspot model, where the background magnetic field is not completely vertical and, thus, the vertical force used as a driver introduces slight transversal perturbations. At these locations there is a very small contribution of waves with magnetic nature, whose behavior is similar to that of fast magneto-acoustic waves.

At the chromosphere ($z=0.8$ Mm) the averaged vertical acoustic flux is $2\times 10^5$ erg cm$^{-2}$s$^{-1}$. The reduction of the acoustic flux with height is caused by several factors, including the radiative losses, the dissipation produced by shocks, and the inability of waves with frequency below the cut-off frequency to propagate to higher layers. The order of magnitude of the average magnetic energy flux at the chromosphere is $10^3$ erg cm$^{-2}$s$^{-1}$.

\begin{figure}[!ht] 
 \centering
 \includegraphics[width=9cm]{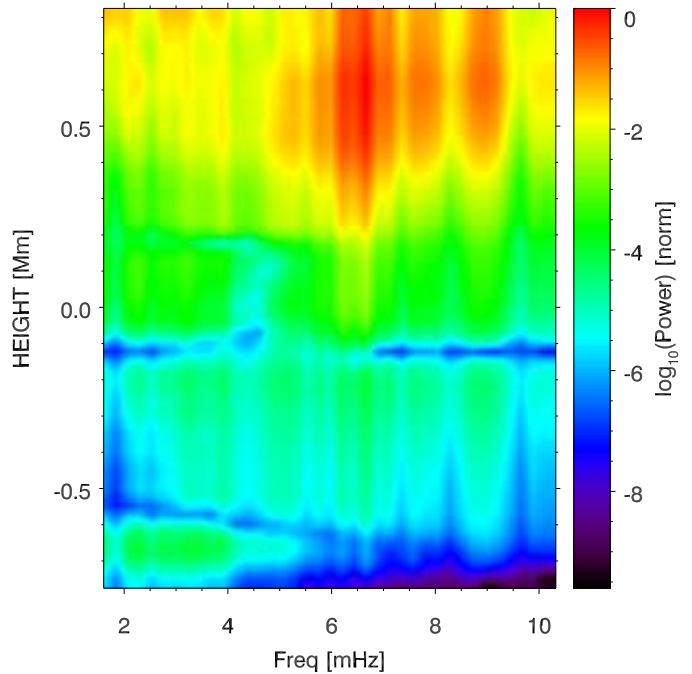}
  \caption{Temperature power spectra in logarithmic scale.}
  \label{fig:power_T}
\end{figure}

Figure \ref{fig:power_T} illustrates the variation of the temperature power spectra with height, spanning from subphotospheric layers to the chromosphere, after averaging for all the horizontal positions inside the umbra. The height with low power at $z=-0.1$ Mm corresponds to the position where the driver was introduced, and it is a consequence of the way the system is driven. As discussed above, the averaged energy flux at that height is zero, and temperature oscillations vanishes for most of the frequencies. Another low power region is found at $-0.5$ Mm for 2 mHz and going down to $-0.6$ Mm for 6 mHz. This minimum in the power of the temperature oscillations may be produced by a node of the evanescent waves. The exact position of the node depends on the height where the driver is introduced and the size of the acoustic cavity, which is different for each frequency. At all heights the frequency of the highest power is between 6 and 7 mHz, as can be seen in Figures \ref{fig:maps_photosphere} and \ref{fig:maps_chromosphere}. The strongest temperature oscillations are found at around $z=0.6$ Mm (0.95 Mm above the umbra surface). In this simulation, radiative transfer was implemented following Newton's cooling law. The photospheric radiative relaxation time $\tau_R$ was obtained from \citet{Spiegel1957} and at $z=0.4$ Mm it takes values as hight as $\tau_R=1877$ s. At the chromosphere we imposed a $\tau_R$ that departs from the the values predicted by Spiegel's formula, since it was derived assuming local thermodynamic equilibrium. We set $\tau_R =10$ s, according to the empirical estimation from \citet{Felipe+etal2010b}. Temperature oscillations are given by an interplay between the amplitude of the waves and the radiative loses. At $z=0.6$ Mm, waves have strong amplitudes and the radiative relaxation time is high, producing strong temperature fluctuations. At higher layers, despite the fact that velocity oscillations are even higher, radiative relaxation time is small and temperature oscillations are damped.

\section{Spectral synthesis}
\label{sect:synthesis}

\subsection{Procedures}
\label{sect:nicole}

For the spectral line synthesis we use the NLTE code NICOLE \citep[][in preparation]{SocasNavarro+etal2014}, which is able to compute
efficiently various combinations of photospheric and chromospheric
spectral lines in simulation datacubes. Stokes polarization induced by
the Zeeman effect is also computed in the presence of magnetic fields.

In NLTE mode, the code assumes complete angle and frequency
redistribution \citep[a good approximation for the \CaII\ infrared triplet, as
shown by][]{Uitenbroek1989} and a 1D plane-parallel
geometry inside the (x,y) column being considered. The \CaII\ model atom
and the line transition atomic parameters are the same as those in \citet{delaCruz-Rodriguez+etal2012}. For the \FeI\ lines we use LTE with the
same atomic parameters as \citet{SocasNavarro2011} and the $log(gf)$
measured by \citet{Bard+etal1991} for the 6301.5 \AA \ transition. In all
cases, collisional broadening is treated with the formalism of \citet{Anstee+O'Mara1995}.


\subsection{\FeI\ 6301.5 \AA\ synthetic observations}
\label{sect:Fe}

The iron lines \FeI\ $\lambda$ 6301.5 and 6302.5 \AA\ are suitable for observing the lower photosphere \citep{delToroIniesta2003}. They are commonly used by several instruments, including the Solar Optical Telescope aboard the HINODE satellite. In this work, we have decided to focus on the \FeI\ $\lambda$ 6301.5 \AA\ because it has a lower Land\'e factor and provides a better result under the weak field approximation (see Section \ref{sect:magnetic field_Fe}).

\begin{figure}[!ht] 
 \centering
 \includegraphics[width=9cm]{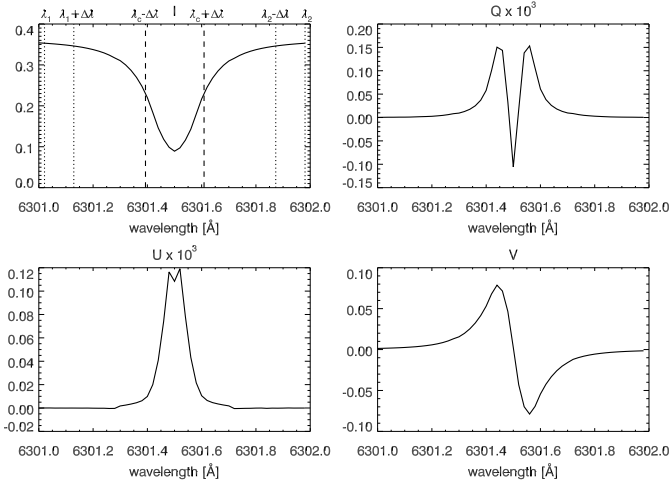}
  \caption{Stokes profiles from \FeI\ 6301.5 \AA\ line at $x=-1.7$ Mm produced by the static background atmosphere. Top left: intensity, top right: Stokes Q, bottom left: Stokes U, bottom right: Stokes V. In top left panel, vertical dashed lines encompass the region used for the integration of the core intensity, and vertical dotted line indicate the region used for the integration of the continuum intensity.}
  \label{fig:Fe_stokes_plot}
\end{figure}

\begin{figure}[!ht] 
 \centering
 \includegraphics[width=9cm]{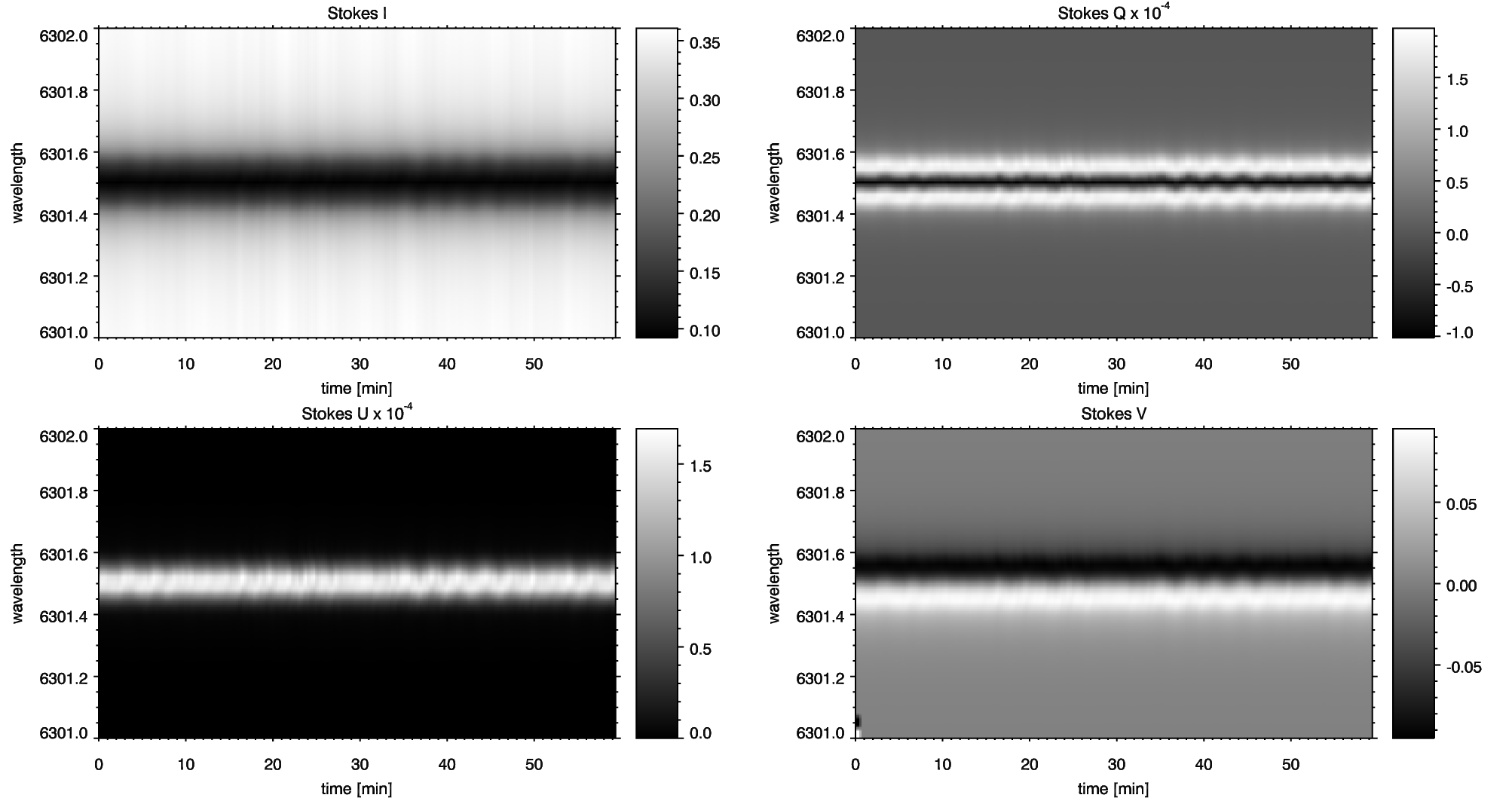}
  \caption{Temporal evolution of the Stokes profiles from \FeI\ 6301.5 \AA\ line at $x=-1.7$ Mm. Top left: intensity, top right: Stokes Q, bottom left: Stokes U, bottom right: Stokes V.}
  \label{fig:Fe_stokes_lambda}
\end{figure}

\begin{figure}[!ht] 
 \centering
 \includegraphics[width=9cm]{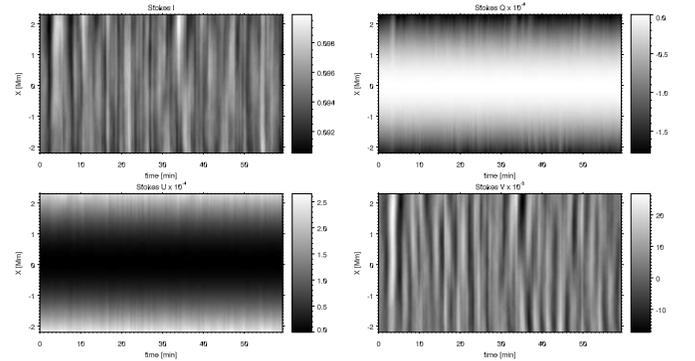}
  \caption{Temporal evolution and spatial variation of the Stokes profiles at the center wavelength from \FeI\ 6301.5 \AA\ line. Top left: intensity, top right: Stokes Q, bottom left: Stokes U, bottom right: Stokes V.}
  \label{fig:Fe_stokes_x}
\end{figure}

Figure \ref{fig:Fe_stokes_plot} shows the synthesized Stokes profiles for the \FeI\ $\lambda$ 6301.5 \AA\ line produced by the MHS atmosphere at 1.7 Mm away from the center of the sunspot. The temporal evolution of the Stokes profiles for this line at that position is plotted in Figure \ref{fig:Fe_stokes_lambda}. The Doppler oscillatory pattern is clearly visible in all the Stokes parameters. Figure \ref{fig:Fe_stokes_x} illustrates the spatial and temporal variation of the Stokes parameters at 6301.5 \AA, \ie, at the center wavelength of the \FeI\ line. Intensity and Stokes V exhibit a wave pattern similar to that previously shown by velocity and density oscillations in Figure \ref{fig:maps_photosphere}. The intensity of the \FeI\ $\lambda$ 6301.5 \AA\ line at the core is modified by Doppler shifts (velocity oscillations) and opacity effects (density oscillations), as will be discussed later. The linear polarization measured from Stokes Q and U vanishes near the center of the umbra, since magnetic field there is mostly vertical, while at farther distances some oscillations are noticeable.  

We have measured the continuum intensity $I_{\rm cont}(x,t)$, the intensity in the line core $I_{\rm core}(x,t)$, and the Doppler velocity obtained from the I and V Stokes profiles. The core intensity is estimated from the integration of the intensity in a wavelength range around the center of the line according to the following expression:

\begin{equation}
I_{\rm core}(x,t)=\int_{\lambda_c-\Delta\lambda}^{\lambda_c+\Delta\lambda} I(\lambda,x,t) {\rm d}\lambda,
\end{equation}

\noindent where $I(\lambda,x,t)$ is the synthetic Stokes I profile, $\lambda_c$ is the center wavelength, and $\Delta\lambda$ is 0.108 \AA. The continuum intensity is retrieved as

\begin{equation}
I_{\rm cont}(x,t)=\int_{\lambda_1}^{\lambda_1+\Delta\lambda} I(\lambda,x,t) {\rm d}\lambda + \int_{\lambda_2-\Delta\lambda}^{\lambda_2} I(\lambda,x,t) {\rm d}\lambda,
\end{equation}

\noindent where $\lambda_1=\lambda_c-0.48$ \AA\ and  $\lambda_2=\lambda_c+0.48$ \AA. The limits for the integration regions are shown in Figure \ref{fig:Fe_stokes_plot}. The intensity fluctuation in the line core and continuum are defined as

\begin{equation}
\delta I_{\rm core}(x,t)=\big(I_{\rm core}(x,t)-I_{\rm core}(x,0)\big)/I_{\rm core}(x,0),
\end{equation}

\begin{equation}
\delta I_{\rm cont}(x,t)=\big(I_{\rm cont}(x,t)-I_{\rm cont}(x,0)\big)/I_{\rm cont}(x,0),
\end{equation}

\noindent respectively. At the initial time the fluctuations produced by the photospheric driver have not been introduced, so the intensity at $t=0$ corresponds to the intensity generated by the static background atmosphere.  

\begin{figure*}[!ht] 
 \centering
 \includegraphics[width=14cm]{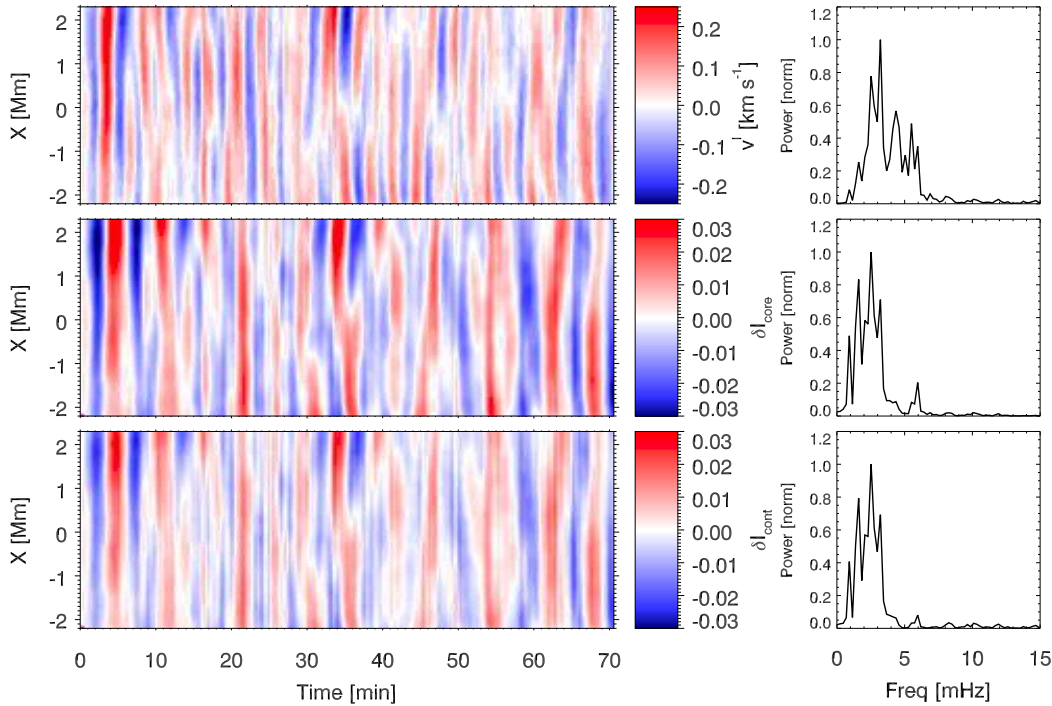}
  \caption{Measurements obtained from the \FeI\ 6301.5 \AA\ line, from top to bottom: Doppler velocity derived from intensity minimum, intensity fluctuations in the core of the line, and continuum intensity fluctuations. Left panels: spatial-velocity maps. Right panels: normalized power spectra. }
  \label{fig:intensity_velocity_fe}
\end{figure*}

Two different Doppler velocities were inferred by measuring the shifts in the wavelength of the intensity minimum and the Stokes V zero cross-position. Both measurements produce very similar spatio-temporal velocity maps. Figure \ref{fig:intensity_velocity_fe} shows the fluctuations in the core intensity and velocity obtained from the Doppler shift of the intensity minimum $v^I$. A comparison of the Doppler velocity map with the actual velocity oscillatory pattern of the simulation at constant geometrical depth (Figure \ref{fig:maps_photosphere}) reveals a strong agreement. The correlation between these two velocity maps is almost unity. In order to estimate the geometrical height where the velocity response function of the \FeI\ $\lambda$ 6301.5 \AA\ line is maximum, we have measured the rms amplitude of the difference between the Doppler velocity measured from the synthesized spectra and the vertical velocity from the simulation at a certain height $z$. The comparison of the results at several $z$ reveals that the difference in amplitude is lowest at $z=-0.05$ Mm for both the shift of the Stokes V zero crossing and the intensity minimum. Note that the Wilson depression of the sunspot model is 350 km, which means that the formation height is around 300 km above the height where continuum $\tau_{5000}=1$. Despite the fact that the contribution function of the spectral line spans over a broad range of heights, the agreement between the velocity oscillations measured from the Doppler shift of the line and the velocity fluctuations at a single height of the simulations is remarkable. In the following we will focus on the analysis of the intensity minimum Doppler velocity. 

Right panels from Figure \ref{fig:intensity_velocity_fe} illustrate the power spectra of the spectroscopically determined velocity and intensity oscillations. As expected, Doppler shift velocity power show peaks at the same frequencies of the actual velocity from the numerical simulations (top right panel of Figure \ref{fig:maps_photosphere}). On the other hand, the power peaks of both the core and continuum intensity fluctuations are located exactly at the same frequencies of the power peaks from the density (middle right panel of Figure \ref{fig:maps_photosphere}). This fact, together with the low amplitude of the temperature oscillations and their different power distribution point to the fluctuations in the optical depth as the cause of the intensity oscillations, rather than the temperature fluctuations produced by the slow wave propagation.

\subsubsection{Photospheric phase relations}
\label{sect:phase_phot}
The propagation of magnetoacoustic-gravity waves in the magnetized solar atmosphere is described by a set of linearized MHD equations. There is an extensive literature about analytical modeling in the case of plane-parallel, gravitationally stratified isothermal atmospheres permeated by a constant magnetic field \citep[\eg,][]{Ferraro+Plumpton1958, Scheuer+Thomas1981, Zhugzhda+Dzhalilov1984a, Wood1990, Khomenko+etal2003}. The mass conservation can be written as

\begin{equation}
\frac{d\rho_{\rm 1}}{dt}+{\bf v_{\rm 1}}\cdot\nabla \rho_{\rm 0}+\rho_{\rm 0}\nabla\cdot{\bf v_{\rm 1}}=0,
\end{equation}

\noindent where $\rho$ is the density, ${\bf v}$ is the velocity, the subindex $0$ indicates the background value, and the subindex $1$ refers to the variations of the variable from the background state. In a two-dimensional case, we can assume a temporal and spatial dependence of the perturbations of the form ${\rm exp}[i(\omega t+k_{\rm x}x+k_{\rm z}z)]$, where $\omega$ is the frequency, and $k_{\rm x}$ and $k_{\rm z}$ are the horizontal and vertical wavenumbers. The following relation between the oscillations in density and velocity is obtained:

\begin{equation}
\label{eq:rho1_v}
\frac{\rho_{\rm 1}}{\rho_{\rm 0}}=\frac{1}{\omega}\Big[v_{\rm z}(\frac{i}{H_{\rm p}}-k_{\rm z})-k_{\rm x}v_{\rm x})\Big],
\end{equation}

\noindent where $H_{\rm p}=\frac{p_{\rm 0}}{\rho_{\rm 0}g}$ is the pressure scale height and $g$ is the gravity. Following Equation 12 from \citet{Khomenko+etal2003}, the dispersion relation for the slow mode in the case of vertical propagation ($k_{\rm x}=0$) in a gravitationally stratified, plane-parallel, isothermal atmosphere with energy losses due to radiative relaxation under the Newtonian cooling approximation, and vertical magnetic field is given by

\begin{equation}
\label{eq:kz_vertical}
k_{\rm z}=\frac{i}{2H_{\rm p}}\pm\frac{1}{2H_{\rm p}}\sqrt{\frac{\gamma\omega^2}{\gamma^*\omega_{\rm c}^2}-1},
\end{equation}

\noindent where $\gamma$ is the ratio of specific heats, $\omega_{\rm c}^2=\frac{g\rho_{\rm 0}}{2\gamma p_{\rm 0}}$ is the square of the isothermal cut-off frequency, and $\gamma^*=\frac{i\gamma\omega+\tau_R^{-1}}{i\omega+\tau_R^{-1}}$ is a measure of radiative losses, with $\tau_R$ as the radiative relaxation time. 
 
\begin{figure}[!ht] 
 \centering
 \includegraphics[width=9cm]{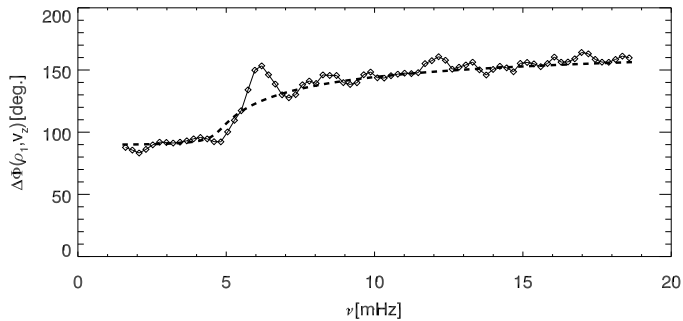}
  \caption{Phase difference between density and vertical velocity at the photosphere. Solid line with diamonds: measured from the simulation, thick dashed line: obtained analytically from Equation \ref{eq:rho1_v}.}
  \label{fig:dfase_rho1_vz}
\end{figure}

The relation between the oscillations in density and vertical velocity produced by a vertical propagating slow mode can be obtained by introducing Equation \ref{eq:kz_vertical} in Equation \ref{eq:rho1_v} and setting $k_{\rm x}=0$. We have computed the phase difference between those two variables at a wide range of frequencies using the density and pressure of the background atmosphere at $z=-0.05$ Mm (the height where the difference in amplitude between the Doppler velocity from the \FeI\ $\lambda$ 6301.5 \AA\ line and the vertical velocity from the simulation is minimum), $\gamma=5/3$, and $\tau_R$ was estimated according to \citet{Spiegel1957}. The result is plotted in Figure \ref{fig:dfase_rho1_vz} and compared with that obtained from the simulation. The latter was measured by averaging the phase shift between density and vertical velocity at all spatial positions inside the umbra at $z=-0.05$ Mm. Positive vertical velocities are directed towards the interior of the Sun, in the direction of the gravity, and a positive phase shift means that the velocity lags the density. The phase shift between density and velocity oscillations from the simulations shows a perfect agreement with the analytical prediction for a slow wave at all frequencies, except around 6 mHz, confirming the nature of the waves driven in the numerical computation.

Observed fluctuations in magnetic field and intensity are produced by intrinsic oscillations, that is, variations of the magnetic field and temperature associated to a propagating wave, and by opacity effects \citep{Lites+etal1998, BellotRubio+etal2000, Khomenko+etal2003, Fujimura+Tsuneta2009}. When a compressible wave travel through the atmosphere, the fluctuations in temperature and density associated with its propagation can generate oscillations in the opacity. These fluctuations move upward and downward the response height of the spectral line and, if the atmosphere has a gradient in magnetic field or temperature, they will produce an apparent fluctuation in magnetic field or intensity, respectively.

In the following we will focus on intensity fluctuations. An increase in the density is associated with an increase in the optical depth and, thus, the formation height of the spectral line is shifted to higher layers. Around the forming region of the \FeI\ $\lambda$ 6301.5 \AA\ line ($z=-0.05$ Mm in our reference system) the temperature of the atmosphere decreases with height. This way, a shift in the formation height to higher layers is accompanied by a reduction of the temperature that the spectral line ``sees'', and vice versa, meaning that the temperature will be displaced by 180$^o$ with respect to the density oscillations. The phase difference between density and vertical velocity oscillations $\Delta\phi (\rho ,v_{\rm z})$ is plotted on Figure \ref{fig:dfase_rho1_vz}. The phase difference between vertical velocity and intensity fluctuations in the line core produced by opacity oscillations is

\begin{eqnarray}
\label{eq:dfase_v_i}
\lefteqn{\Delta\Phi(v_{\rm z}, I_{\rm core})=\Phi(v_{\rm z})-\Phi(I_{\rm core})=}\nonumber\\
&&\Phi(v_{\rm z})-\Phi(\rho)+180^o=\nonumber\\
&&180^o-\Delta\Phi  (\rho ,v_{\rm z})
\end{eqnarray}

\begin{figure}[!ht] 
 \centering
 \includegraphics[width=9cm]{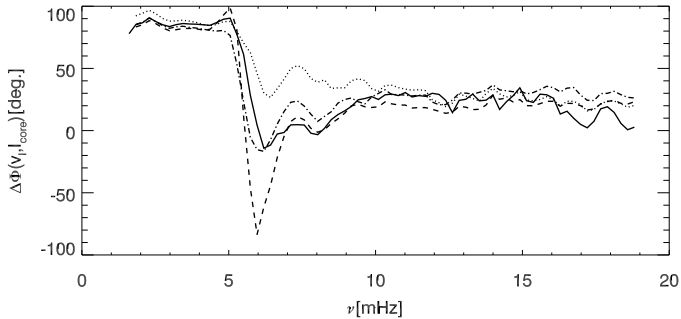}
  \caption{Phase difference between several variables at the photosphere. Solid line: between Doppler velocity and intensity at the core of \FeI\ $\lambda$ 6301.5 \AA\ line obtained from the synthetic observations; dotted line: between vertical velocity from the simulation and intensity obtained from Equation \ref{eq:dfase_v_i}; dashed line: between velocity at $\tau (Fe)$ and $T_{\rm 0}^{\rm \tau (Fe)}$; dashed-dotted line: between velocity at $\tau (\FeI)$ and $T_{\rm 0+1}^{\rm \tau (Fe)}(x,t)$.}
  \label{fig:dfase_vz_I}
\end{figure}

Figure \ref{fig:dfase_vz_I} illustrates the variation of the phase shift between vertical velocity and line core intensity with frequency measured with the \FeI\ $\lambda$ 6301.5 \AA\ line (solid line). Lower frequencies (from 1.5 mHz up to 5 mHz) show a 90$^o$ shift, with the velocity ahead of the intensity. Between 5 and 6 mHz there is an steep decrease in the phase shift, which takes negative values of almost -20$^o$. For higher frequencies velocity oscillations are leading again, with $\Delta\Phi(v_{\rm z}, I_{\rm core})$ around 20$^o$. The phase difference expected from Equation \ref{eq:dfase_v_i} (dotted line) shows a good agreement with the measured phase shift at frequencies below 5 mHz and above 10 mHz. However, in the frequency range between 5 and 10 mHz it produces significant higher phase shifts. 

In order to further explore the nature of the intensity oscillations of the line core, we have computed the phase shift between vertical velocity and temperature at constant optical depth $log(\tau (\FeI))=-1$. The value of $\tau (\FeI)$ was chosen as the average optical depth for all the spatial positions inside the umbra at $z=-0.05$ Mm, that is, at the height where the velocity response function is maximum. We have computed the time-dependent optical depth and constructed velocity and temperature spatio-temporal maps by interpolating their values to the height where $log(\tau (\FeI))=-1$. Two different temperature maps were obtained. In the first case, we took the temperature from the atmospheric MHS model, retrieving the temperature maps $T_{\rm 0}^{\rm \tau (Fe)}(x,t)$. In the second case, we used the total temperature, including the background plus the temperature oscillations associated to the propagation of the magnetoacoustic waves. The later will be referred as $T_{\rm 0+1}^{\rm \tau (Fe)}(x,t)$. 

The phase shift between velocity and $T_{\rm 0}^{\rm \tau (Fe)}(x,t)$ reproduce the phase difference between velocity and line core intensity for all frequencies but the range between 5 and 6.5 mHz, where it shows a much smaller phase shift, in the sense that at those frequencies the velocity lags the temperature. On the other hand, by introducing the intrinsic oscillations of the temperature (associated to wave propagation), the agreement is significantly better, as can be seen from the phase shift between the velocity and $T_{\rm 0+1}^{\rm \tau (Fe)}(x,t)$ (dashed-dotted line). In this case, the phase shift around 6 mHz shows a remarkable agreement with that measured between Doppler velocity and intensity at the core of \FeI\ $\lambda$ 6301.5 \AA\ line. Note that these frequencies correspond to the power peak in the photospheric temperature oscillations (bottom right panel from Figure \ref{fig:maps_photosphere}), and point out that for upward propagating slow waves in a sunspot umbra, photospheric intensity oscillations in the \FeI\ $\lambda$ 6301.5 \AA\ line are mainly produced by opacity effects, except for frequencies between 5 and 6.5 mHz, where the intrinsic temperature oscillations are more prominent and affect the intensity of the line core.

\subsubsection{Measuring photospheric magnetic field}
\label{sect:magnetic field_Fe}

We have inferred the magnetic flux density oscillations using the weak field approximation \citep[\eg.][]{Landi-Degl'Innocenti1992} for the \FeI\ $\lambda$ 6301.5 \AA\ line. In this limit, the Stokes I and V profiles are related as

 \begin{equation}
\label{eq:weak_field}
V(\lambda)=-\phi C\frac{\partial I(\lambda)}{\partial\lambda},
\end{equation}

\noindent where $\phi =fB\cos\gamma$ is the longitudinal magnetic flux density, $B$ is the field strength, $\gamma$ is the magnetic field inclination, and $f$ is the filling factor. The constant $C=4.6686\times10^{-13}\lambda_0^2\bar{g}$ depends on the spectral transition through its central wavelength $\lambda_0$ (expressed in \AA\ ) and effective Land\'e factor $\bar{g}$. Following \citet{Martinez-Gonzalez+Bellot-Rubio2009}, we have retrieved the longitudinal magnetic flux density $\phi$ from a least-squares minimization. For each spatial position and time step, it is obtained as 

\begin{equation}
\label{eq:phi_weak_field}
\phi =-\frac{\sum_i\frac{\partial I}{\partial\lambda_i}V_i}{C\sum_i(\frac{\partial I}{\partial\lambda_i})^2},
\end{equation}

\noindent where the index $i$ samples all the wavelengths across the profile. 

\begin{figure*}[!ht] 
 \centering
 \includegraphics[width=14cm]{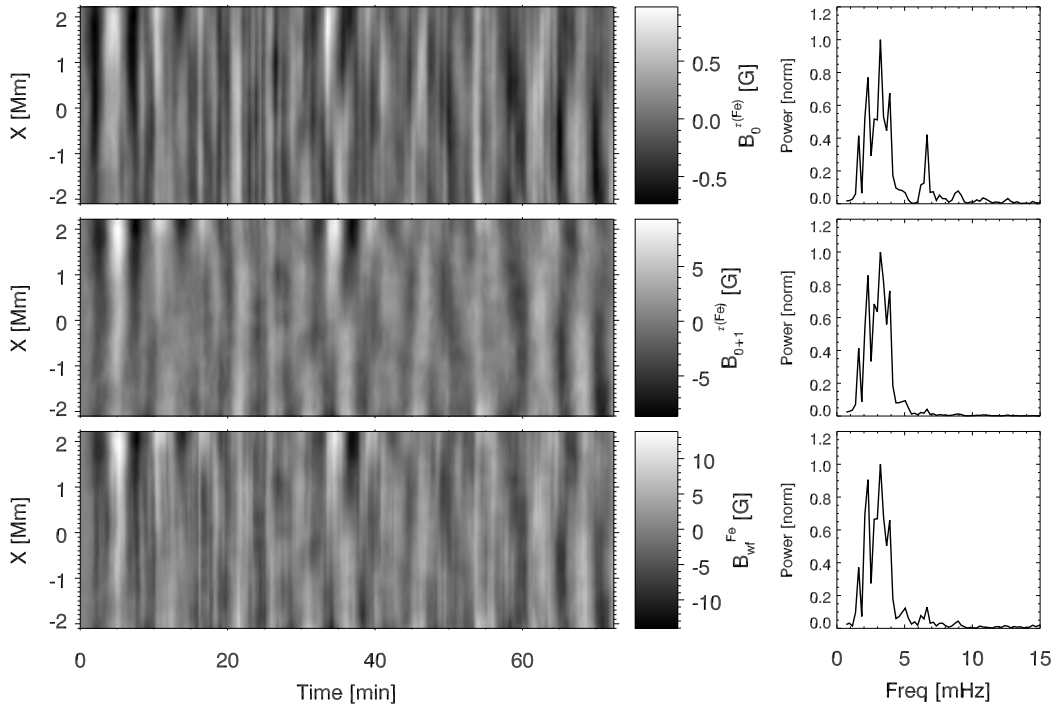}
  \caption{Magnetic field spatio-temporal maps at the photosphere (left panels) and power spectra (right panels). Top panels: magnetic field fluctuations produced by opacity oscillations on the background static magnetic field, middle panels: magnetic field fluctuations at constant optical depth $\tau(Fe)$, bottom panels: magnetic field oscillations retrieved from the \FeI\ $\lambda$ 6301.5 \AA\ line using the weak field approximation.}
  \label{fig:magnetic_field_fe}
\end{figure*}

The atmosphere at the center of the simulated umbra is permeated by the magnetic field at all spatial positions and, thus, $f=1$. In this case, the longitudinal magnetic flux density provides the longitudinal magnetic field strength. Moreover, the magnetic field is almost vertical at all the spatial positions considered in this work, and we can assume $\cos\gamma\approx 1$. This way, from Expression \ref{eq:phi_weak_field} we directly obtain the vertical magnetic field. The inferred fluctuations of the magnetic field from the \FeI\ $\lambda$ 6301.5 \AA\ line are plotted in the bottom panels of Figure \ref{fig:magnetic_field_fe}. The oscillations are concentrated in the 5 minutes band, with the same power peaks previously shown by velocity, density, and intensity oscillations. The strongest oscillations of the inferred magnetic field are found during the first 10 minutes of simulations at 2 Mm far from the axis of the sunspot. Their amplitude are up to 14 G, while the rms amplitude of the umbra including all time steps is 3.44 G.    

We have computed the variations of the magnetic field at constant optical depth in the same way temperature fluctuations at $\tau (\FeI)$ were obtained in the previous section. We also computed two magnetic field maps at constant $\tau$, one of them including only the magnetic field from the static background model ($B_0^{\tau (Fe)}$) in the calculations, and the other using the total magnetic field (with its perturbations in addition to the background magnetic field, $B_{0+1}^{\tau (Fe)}$). The maps are plotted in top and middle panels of Figure \ref{fig:magnetic_field_fe}, respectively. The fluctuations produced purely by opacity oscillations show amplitudes below 1 G, with an rms amplitude of 0.25 G. Most of their power is also at the 5 minutes band, but it has a significant peak at 6.6 mHz, above the cut-off value. Around the \FeI\ $\lambda$ 6301.5 \AA\ line formation height, the gradient $dB/dz$ of our sunspot model is $-0.1$ G km$^{-1}$. This value is significantly lower than several estimations of the magnetic field gradients in sunspot umbrae. \citet{BellotRubio+etal2000} inferred a gradient of about $-3.8$ G km$^{-1}$, while \citet{Westendorp-Plaza+etal2001} found $dB/dz\approx -1.5$ G km$^{-1}$. \citet{Collados+etal1994} compared the differences between the umbra of large and small sunspots, and found an order of magnitude difference between the gradients of the two spots. In the region log$\tau$=[-1.5,-2.0], the vertical gradient of the magnetic field of the small spot is $-2.1$ G km$^{-1}$, but for the large sunspot it is $-0.25$ G km$^{-1}$. The later is closer to our model. Note that a magnetic field stratification similar to that inferred for small spots would produce strikingly higher magnetic field fluctuations due to opacity effects.  

Middle panel of Figure \ref{fig:magnetic_field_fe} shows the total oscillations of the magnetic field at constant optical depth. The wave pattern and order of magnitude of the amplitude is similar to those obtained using the weak field approximation with the \FeI\ $\lambda$ 6301.5 \AA\ line. The highest amplitudes are below 10 G and the rms amplitude is 2.1 G. The weak field calculation overestimates the amplitude of magnetic field oscillations by a factor around 1.6.

\begin{figure*}[!ht] 
 \centering
 \includegraphics[width=14cm]{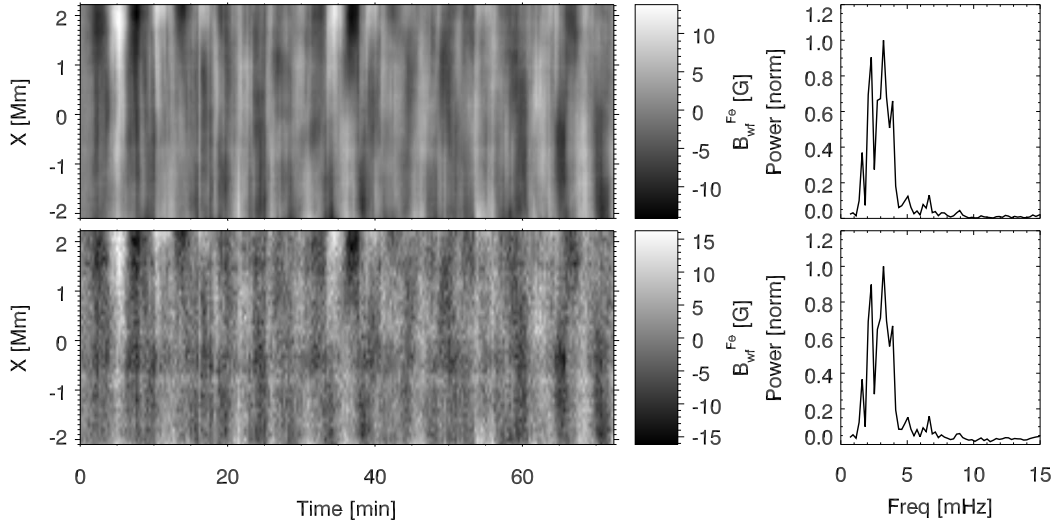}
  \caption{Magnetic field spatio-temporal maps at the photosphere (left panels) and power spectra (right panels) obtained from the \FeI\ $\lambda$ 6301.5 \AA\ line using the weak field approximation. Top panels: adding a random noise of $10^{-4}I_{\rm cont}$, bottom panels: adding a random noise of $10^{-3}I_{\rm cont}$.}
  \label{fig:magnetic_field_fe_noise}
\end{figure*}

The previous estimation of the magnetic field from the weak approximation was obtained in the absence of noise. As a next step, we have added noise to the Stokes profiles prior to compute the magnetic field using Equation \ref{eq:phi_weak_field}. The calculations were performed with two different noise levels. In the first case, we add a random noise of $10^{-4}I_{\rm cont}$. In the second case, the noise is one order of magnitude higher, $10^{-3}I_{\rm cont}$. The magnetic field inferred from the weak approximation in these two cases is illustrated in Figure \ref{fig:magnetic_field_fe_noise}. The case with $10^{-4}I_{\rm cont}$ noise shows no significant difference with the magnetic field estimated without the inclusion of noise (bottom panel of Figure \ref{fig:magnetic_field_fe}). On the other hand, the higher noise case shows a noisier magnetic field pattern. However, magnetic field perturbations in the range 5-10 G remain above the noise level.

\subsection{\CaII\ IR lines synthetic observations}
\label{sect:Ca}
\begin{figure}[!ht] 
 \centering
 \includegraphics[width=9cm]{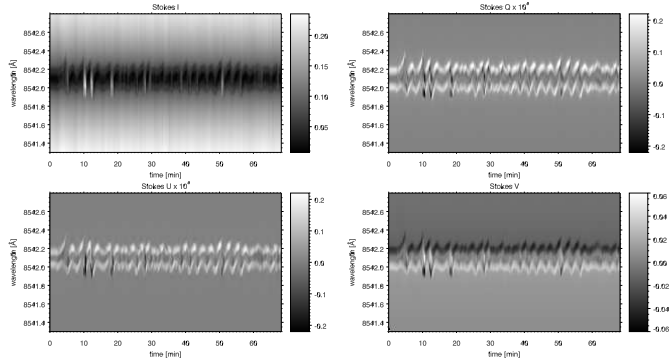}
  \caption{Temporal evolution of the Stokes profiles from \CaII\ 8542.09 \AA\ line at $x=1.1$ Mm. Top left: intensity, top right: Stokes Q, bottom left: Stokes U, bottom right: Stokes V.}
  \label{fig:Ca_stokes_lambda}
\end{figure}

\begin{figure}[!ht] 
 \centering
 \includegraphics[width=9cm]{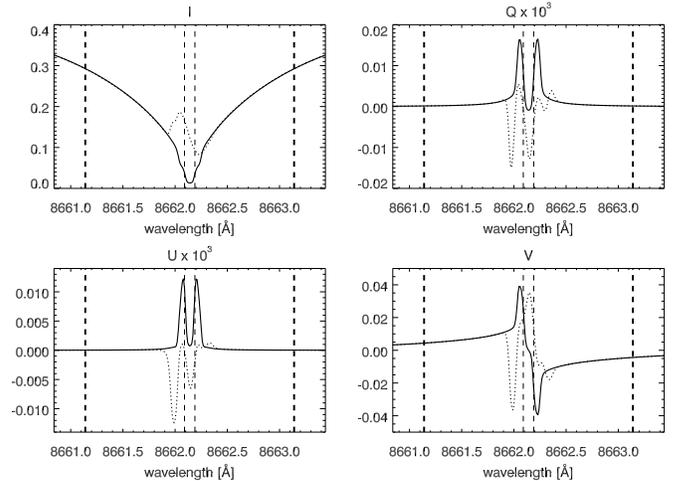}
  \caption{Stokes profiles from \CaII\ 8542.09 \AA\ line at $x=1.1$ Mm produced by the static background atmosphere (solid line) and during a shock at $t=744$ s (dotted line). Top left: intensity, top right: Stokes Q, bottom left: Stokes U, bottom right: Stokes V. In top left panel, vertical dashed lines encompass the region used for the integration of the core intensity, and vertical dotted line indicate the region used for the integration of the continuum intensity. Vertical dashed lines indicate the regions used for the measurement of the magnetic field in Section \ref{sect:magnetic_field_Ca}}
  \label{fig:Ca_stokes_plot}
\end{figure}

The temporal evolution of Stokes I, Q, U, and V for the chromospheric \CaII\ $\lambda$ 8542.09 \AA\ line at $x=1.1$ Mm position is plotted in Figure \ref{fig:Ca_stokes_lambda}. Similar variation of the Stokes profiles are found for the  \CaII\ IR lines at 8498.02 and 8662.14 \AA, not shown in the figure. We have chosen to show one of the locations which presents higher amplitude waves at the chromosphere, as can be seen in Figure \ref{fig:maps_chromosphere}. All the Stokes profiles show strong Doppler shifts, which consist on a progressive displacement of the spectral line towards the red (higher wavelengths) followed by a sudden shift towards the blue (lower wavelengths). It is interesting to note that time steps with strongest shifts, when shock waves reach the chromosphere, are accompanied by highly abnormal Stokes profiles. They are clearly seen at $t=10, 12, 18, 50$ min. Figure \ref{fig:Ca_stokes_plot} shows a comparison between the Stokes profiles of the \CaII\ $\lambda$ 8542.09 \AA\ line at rest and those produced during a shock. In the case of the later, intensity shows striking emission self-reversal shifted to the blue side of the line. For Stokes V, during the shocks the profile departs from the normal state (composed by two lobes with the same polarity and, thus, opposite sign) to form a highly asymmetric profile. While the normal profiles show a positive value for the blue lobe and a negative value for the red lobe, the incidence of a shock produces negative values of Stokes V in the blue side of the spectral line and positive around the core and the red side of the line. The effects of the shocks are also evident in the linear polarization Stokes profiles Q and U. At $x=1.1$ Mm, for the static background atmosphere both Q and U show symmetric profiles with two positive lobes. During the shocks, they are significantly modified, with negative values which usually start at the blue side of the line and later they appear at the core and at the red lobe. The influence of the shocks on the Stokes profiles will be discussed in detail in Section \ref{sect:shocks}.       

\begin{figure}[!ht] 
 \centering
 \includegraphics[width=9cm]{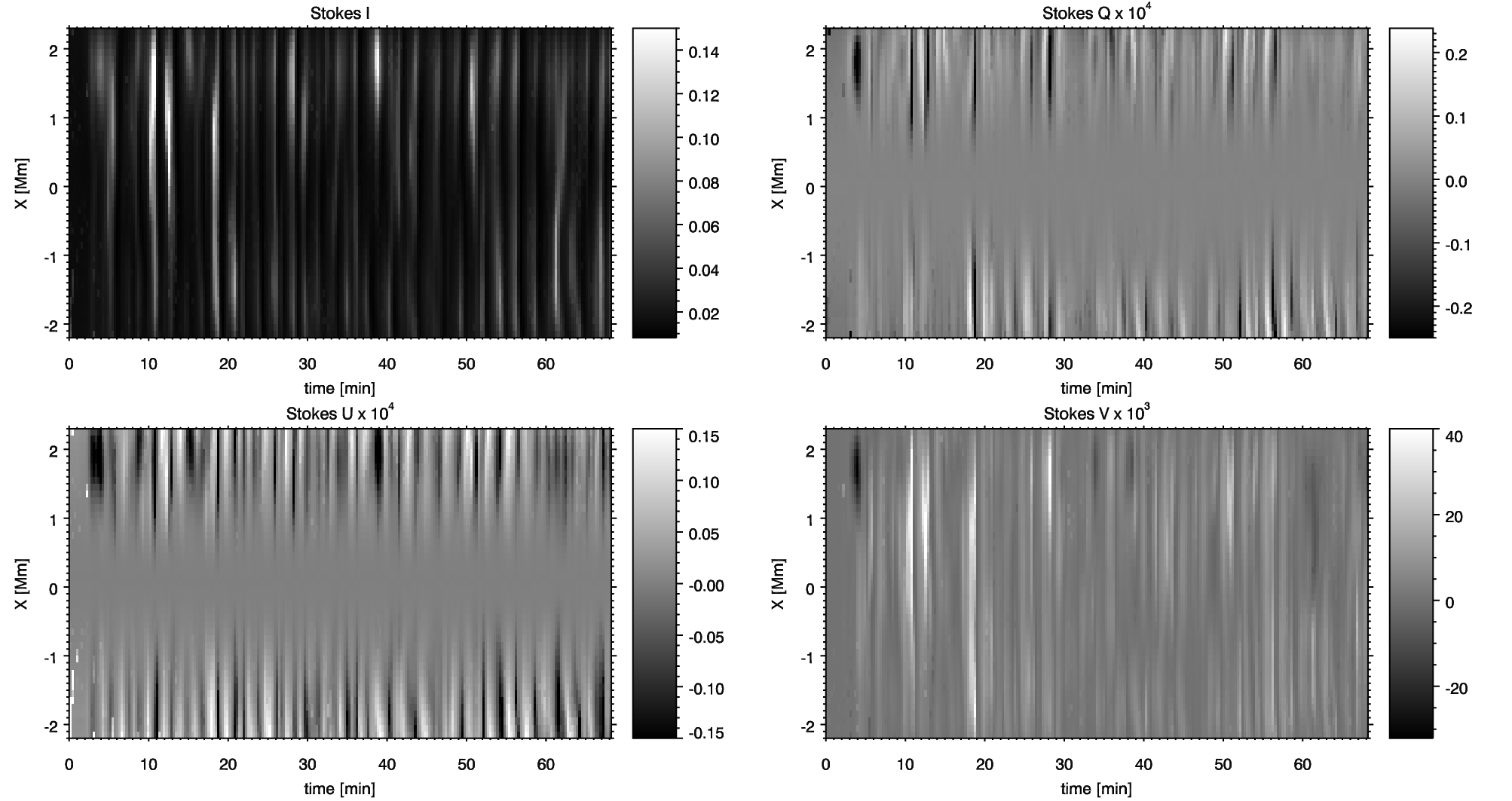}
  \caption{Temporal evolution and spatial variation of the Stokes profiles at the center wavelength from \CaII\ 8542.09 \AA\ line. Top left: intensity, top right: Stokes Q, bottom left: Stokes U, bottom right: Stokes V.}
  \label{fig:Ca_stokes_x}
\end{figure}

Figure \ref{fig:Ca_stokes_x} shows some spatio-temporal maps of the Stokes profiles at 8542.09 \AA\, that is, at the wavelength of the core of the \CaII\ line at rest. The wave pattern of the high frequency waves that reach the chromosphere is visible in all the Stokes parameters, and it is strongly affected by the abnormal profiles produced by the incidence of shocks. At those time steps, the intensity and Stokes V are greatly enhanced. As expected, Stokes Q and U oscillations are only visible out of the axis of the sunspot, since they vanish for vertical magnetic fields.

\subsubsection{Chromospheric velocity from \CaII\ IR lines}

\begin{figure*}[!ht] 
 \centering
 \includegraphics[width=14cm]{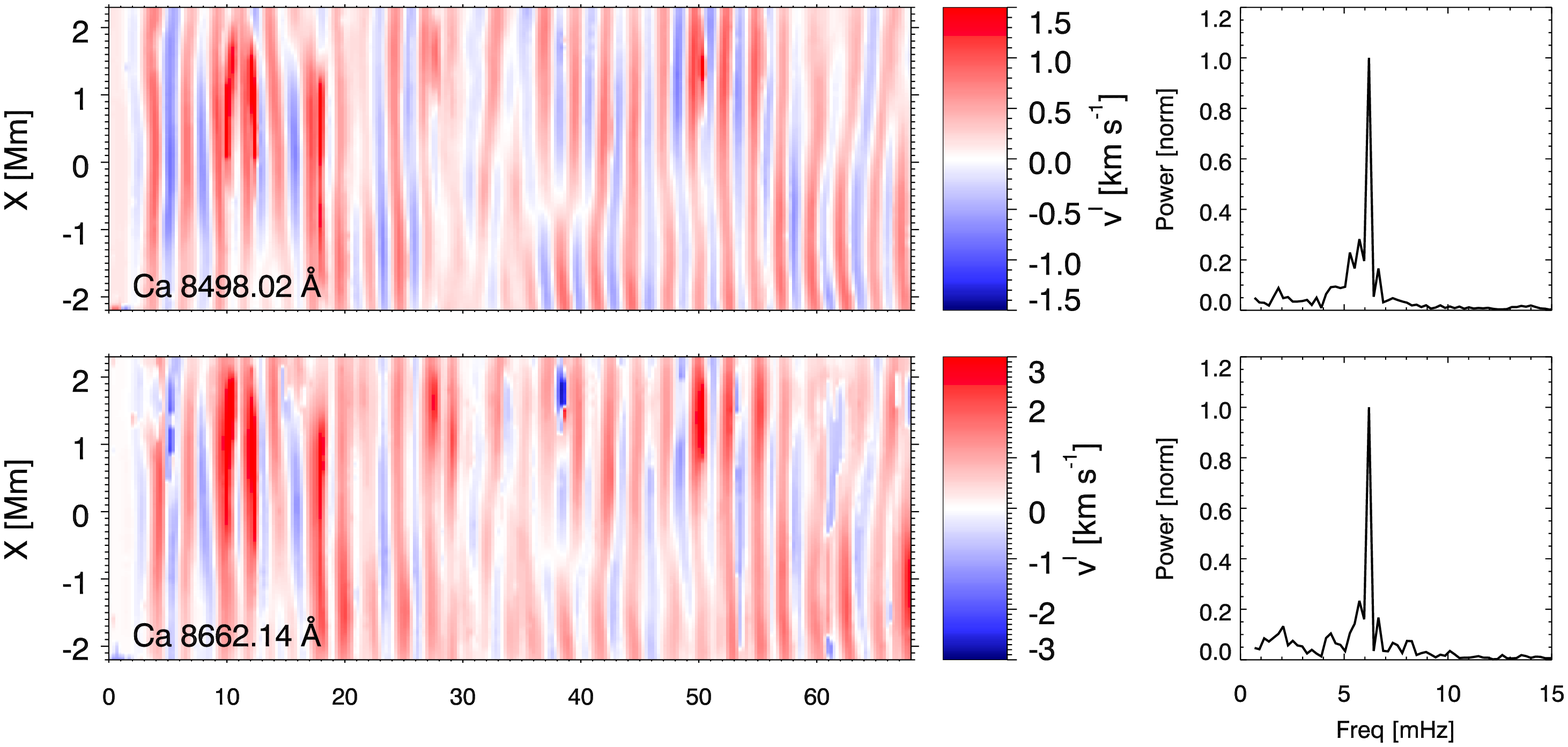}
  \caption{Spatio-temporal maps of the Doppler velocity measured with the \CaII\ 8498.02  \AA\ (top panel) and 8662.14 \AA\ (bottom panel) lines. Right panels show their power spectra.}
  \label{fig:velocity_ca}
\end{figure*}

The Doppler velocity retrieved from the position of the minimum of the \CaII\ $\lambda$ 8498.02 and 8662.14 \AA\ lines is shown in Figure \ref{fig:velocity_ca}. The \CaII\ $\lambda$ 8542.09 \AA\ line (not plotted) shows a similar behavior. Both maps show a preponderance of redshifts. Note that we have defined the position of the intensity minimum as the reference wavelength for the Doppler shift even for the highly abnormal intensity profiles during shocks (dotted line in Figure \ref{fig:Ca_stokes_plot}). At those time steps it may not be a reliable proxy for the position of the line core. A comparison of the Doppler maps with the chromospheric vertical velocity obtained directly from the simulation (Figure \ref{fig:maps_chromosphere}) reveals several similitudes and some differences. In both cases the power is concentrated at a single high frequency peak. The frequency of the vertical velocity peak from the simulation is 6.6 mHz, while that obtained from the Doppler shift is 6.2 mHz. Most of the redshifts measured from the Doppler shift of the spectral lines can be identified with their corresponding positive vertical velocity in the simulation. However, for some of the negative velocities, especially those with higher amplitudes, the Doppler signal does not show a blueshift (\ie, at $t=10$ and $t=12$ min). As shown previously, when a shock reaches the chromosphere a emission self-reversal is produced in the intensity of the \CaII\ lines. In those cases, the position of the intensity minimum does not provide a reliable wavelength for the core of the line, since it is shifted towards the red, and the real velocity oscillatory signal is concealed by the effects of the shocks on the intensity profiles. On the other hand, the highly asymmetrical Stokes V profiles generated by the shocks also impedes us to use the zero-crossing wavelength as a proxy for measuring the Doppler shift.

\subsubsection{Effect of shocks on the Stokes profiles}
\label{sect:shocks}
High frequency (above the cut-off) slow acoustic waves can propagate from the photosphere to the chromosphere. As they reach higher layers with lower density, their amplitude increase in order to conserve the energy. At a certain height, some wavefronts can steepen into shocks, causing dramatic changes in the atmosphere. The upward propagating shock compresses the plasma in front of it and produces an increase of the temperature, with the corresponding changes in optical depth and response height of the chromospheric spectral lines. The combination of these effects with the strong velocity gradients associated with the shocks and the vertical gradient of the atmospheric magnetic field leads to striking modifications of the Stokes profiles.

\begin{figure*}[!ht] 
 \centering
 \includegraphics[width=18cm]{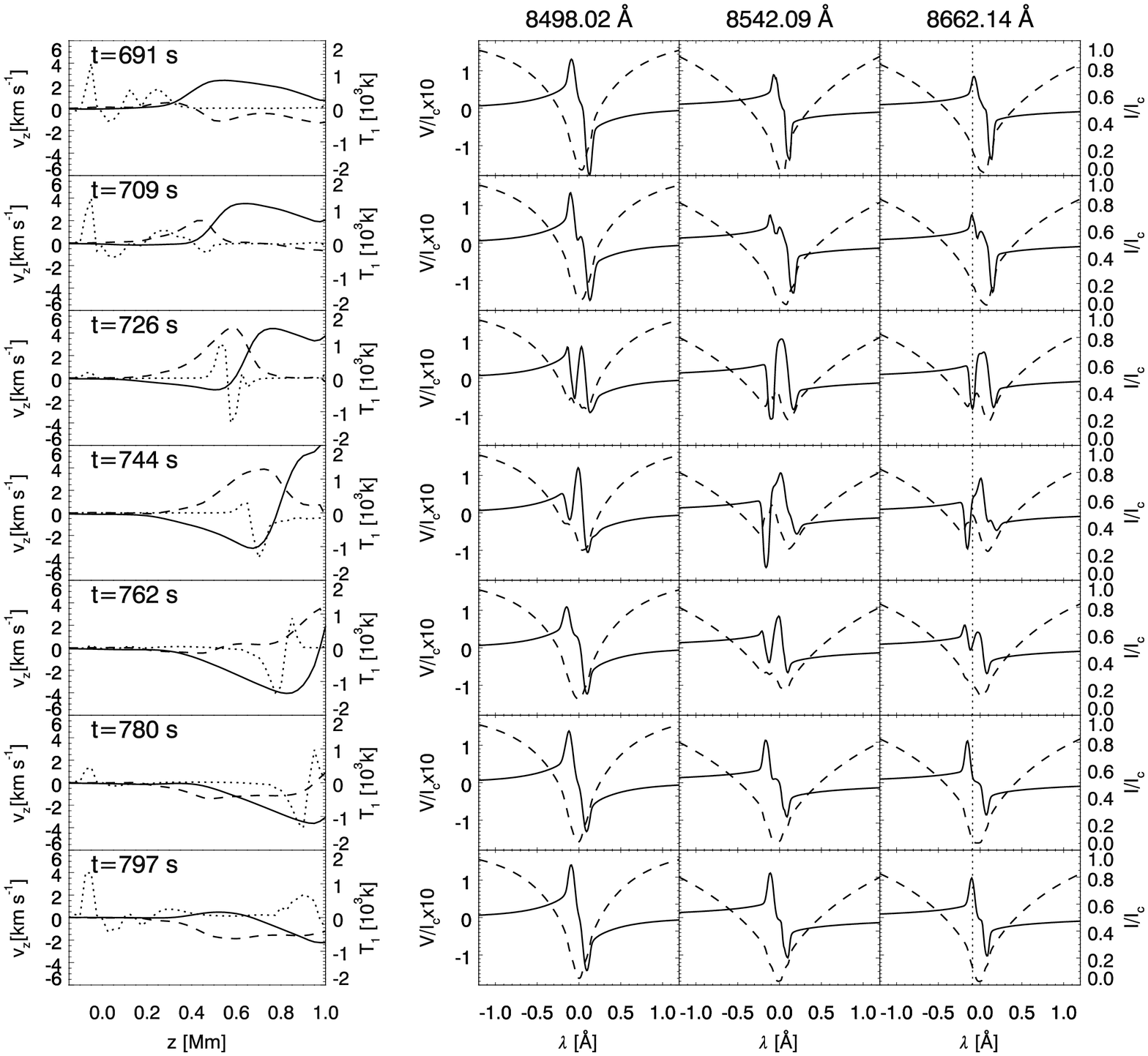}
  \caption{Temporal evolution of a shock. Each row corresponds to a time step between 691 and 797 after the beginning of the simulation. Left panels: variation of the vertical velocity (left ordinate, solid line), perturbation of the temperature (right ordinate, dashed line), and Stokes V response function to temperature at $\lambda=8662.05$ \AA\ (normalized, dotted line) with height. Right panels: Stokes V (left ordinate, solid line) and Stokes I (right ordinate, dashed line) profile for the \CaII\ 8498.02, 8542.09, 8662.14 \AA\ lines. The vertical dotted line at the right panels represents the wavelength of the response function plotted in the left panels.}
  \label{fig:shock}
\end{figure*}

Figure \ref{fig:shock} illustrates the temporal evolution of a shock, including the stratification of the vertical velocity and the temperature perturbation of the atmosphere, the Stokes V response functions, and the Stokes I and V profiles of the three \CaII\ lines at each time step between the arrival of the wavefront to the line forming region and the time when the shock has passed through to higher layers. We will start discussing the variation of the intensity. At $t=691$ s the intensity of the three spectral lines is approximately at the rest state, with a smooth shift towards the red according to the positive vertical velocity of the wavefront. At the next time step, the velocity increases up to around 4 km s$^{-1}$ and the redshift of the lines is more prominent, especially for the 8542.09 and 8662.14 \AA\ lines. At $t=726$ s the atmosphere shows a strong velocity gradient, where the velocity in the higher layers is around 4 km s$^{-1}$ and it suddenly drops to negative values in the lower chromosphere. While the core of the line ``sees'' a positive velocity and is shifted towards the red, part of the wings are formed at a lower height with negative velocity and, thus, shifted towards the blue. The shock wave produces a strong increase of the temperature around 1500 k at this lower height, which enhances the intensity emission in the blueshifted wing of the lines. This process generates the self-reversal of the intensity that is clearly visible at time steps 726 and 744 s. It is particularly prominent for the \CaII\ $\lambda$ 8542.09 and 8662.14 \AA\ lines, since their response function is located higher in the atmosphere, but it can be also noticed in the \CaII\ 8498.02 line. As the shock wavefront pass through the formation height of the lines, the emission self-reversal progressively decreases (at $t=762$ s) and finally the intensity returns to the normal profile with a blueshift at $t=780$ s.

The evolution of the Stokes V during a shock shows highly asymmetrical profiles. As seen in the three \CaII\ lines from Figure \ref{fig:shock}, before the shock develops the Stokes V signal presents the usual configuration, showing two lobes with opposite signs. As the shock propagates to the chromosphere, the strengths of the two lobes are progressively reduced, while at the center of the line the Stokes V signal increases. In the case of the blue lobes the variation is more striking, since it changes from positive values at $t=691$ s to a remarkable negative signal at $t=726$ s. When the shock completes its travel through the chromosphere, the Stokes V profiles recover their original shape. In order to discuss in detail the influence of the shocks on Stokes V, we are going to focus on the blue lobe of the \CaII\ $\lambda$ 8662.14 \AA, where some of the most dramatic changes are produced. The dotted line in the left panels of Figure \ref{fig:shock} show the Stokes V response function to temperature at $\lambda=8662.05$ \AA, that is, shifted 0.09 \AA\ to the blue from the core of the \CaII\ $\lambda$ 8662.14 \AA\ line. At each time step the response function has been normalized, and the results are overplotted independently of the scales shown in the left and right ordinates, which refer to velocity and temperature perturbations, respectively. When the amplitude of the chromospheric oscillations is low ($t=691$ s) the main contribution to Stokes V comes from the photosphere, at $z=-0.05$ Mm (0.30 Mm above $\tau_{5000}=1$). However, as the shock propagates to higher layers and its amplitude increases, a new component appears in the response function. This contribution to the Stokes V signal is dominant at $t=726$ s. It shows a complex behavior, with a mixture of positive and negative values depending on the height. A negative Stokes V response function produces the reduction of the blue lobe, reaching negative values in Stokes V as seen at $t=726$ and $744$ s. The chromospheric contribution to the blue lobe of the \CaII\ line follows the propagation of the shock up to $z\approx0.9$ Mm ($t=780$ s) and then it decreases. At this time step, the photospheric response function is comparable to that associated to the shock, and the Stokes V profiles recovers their steady shape. When the wave amplitude is again low, Stokes V signal is formed mainly at the photosphere ($t=797$ s).

\subsubsection{Chromospheric phase relations}
\label{sect:phase_chrom}

\begin{figure}[!ht] 
 \centering
 \includegraphics[width=9cm]{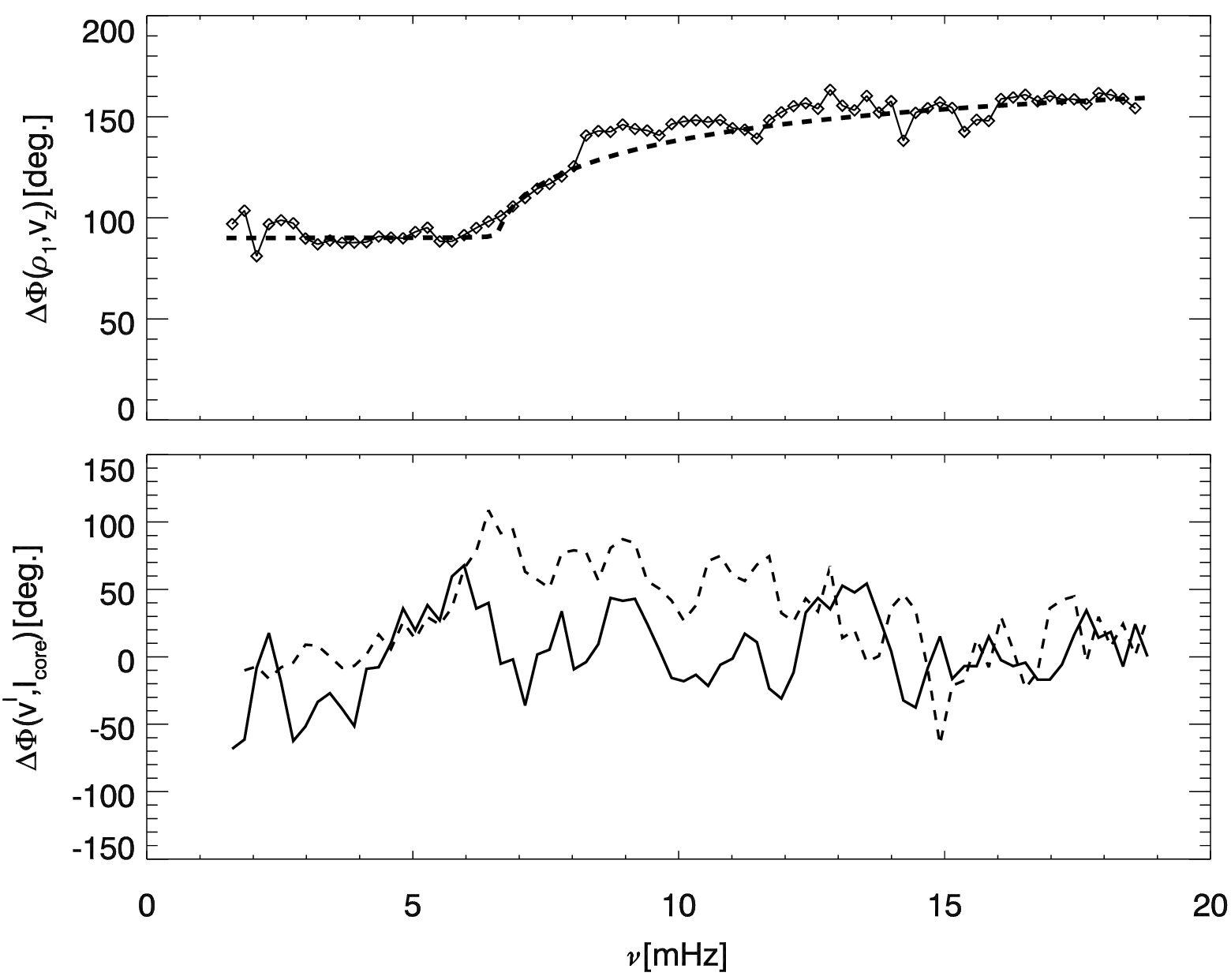}
  \caption{Top panel: phase difference between density and vertical velocity at the chromosphere measured from the simulation (solid line with diamonds) and obtained analytically from Equation \ref{eq:rho1_v} (dashed line). Bottom panel: phase difference between Doppler velocity and intensity at the core of \CaII\ $\lambda$ 8542.09 \AA\ line obtained from the synthetic observations (solid line) and between velocity at $\tau (\CaII)$ and $T_{\rm 0+1}^{\rm \tau (Ca)}(x,t)$ (dashed line).}
  \label{fig:phase_chrom}
\end{figure}

Following the same approach described in Section \ref{sect:phase_phot}, we have evaluated the phase difference between vertical velocity and density at the chromosphere. The solid line with diamonds in the top panel of Figure \ref{fig:phase_chrom} shows the phase difference measured from the simulation at $z=0.5$ Mm (0.85 Mm above the photospheric height in the umbra due to the Wilson depression). Its variation with frequency is similar to that obtained at the photosphere (Figure \ref{fig:dfase_vz_I}). For low frequency waves the phase difference is $90^{o}$, and it increases with the frequency up to $160^{o}$. However, note that at the chromosphere the frequency at which the phase difference starts to increase is higher. While for the photospheric case it is at 4.5 mHz, in the chomosphere it is around 6 mHz. The dashed line shows the analytical phase difference computed from Equation \ref{eq:rho1_v}. The agreement with that measured from the simulation is remarkable.  

Previously, we have shown that at the photosphere, for those frequencies where the temperature oscillations are low, Equation \ref{eq:dfase_v_i} provides a good approximation of the phase difference between oscillations in the vertical velocity and the core intensity. This simple model does not capture the phase relation between those variables at the chromosphere. At those heights wave propagation is accompanied with strong temperature oscillations (Table \ref{tab:simulations} and Figure \ref{fig:maps_chromosphere}), and the assumption that intensity fluctuations are mainly produced by opacity effects is inadequate. In addition, chromospheric oscillations are certainly nonlinear. This is clearly seen in Table \ref{tab:simulations}, which shows than density perturbation is larger than the background density. Thus, the analysis of Section \ref{sect:phase_phot} based on linearized MHD equations cannot capture the full complexity of wave propagation at the chromosphere. Bottom panel of Figure \ref{fig:phase_chrom} illustrates the phase difference between vertical velocity and core intensity measured from the \CaII\ $\lambda$ 8542.09 \AA\ line (solid line). It shows a complex behavior, with sudden variations with frequency in the range between $-50^o$ and $50^o$. 

In a similar fashion to Section  \ref{sect:phase_phot}, we have calculated the phase shift between vertical velocity and temperature at constant optical depth $log(\tau (\CaII))=-5.9$. The value of $\tau (\CaII)$ was selected by looking for the optical depth where the intensity response function at the core of the \CaII\ $\lambda$ 8542.09 \AA\ line is maximum. The dashed line in the bottom panel of Figure \ref{fig:phase_chrom} shows the phase shift between vertical velocity and $T_{0+1}^{\rm \tau(Ca)}$. Following the same notation previously defined, $T_{0+1}^{\rm \tau(Ca)}$ consist in a spatio-temporal map of the total temperature (background atmosphere plus perturbation) at constant optical depth $\tau (\CaII)$. A comparison with the phase shift measured from the Doppler velocity and the core intensity of the \CaII\ $\lambda$ 8542.09 \AA\ reveals some similarities but also significant differences, especially between 6 and 12 mHz. Note that our estimation of the Doppler velocity and core intensity of the \CaII\ IR lines is strongly contaminated by the abnormal intensity profiles produced by the shock waves (Figure \ref{fig:shock}). In this context, the disagreement between both measurements comes at no surprise.

\subsubsection{Measuring magnetic field from \CaII\ IR lines}
\label{sect:magnetic_field_Ca}

\begin{figure*}[!ht] 
 \centering
 \includegraphics[width=14cm]{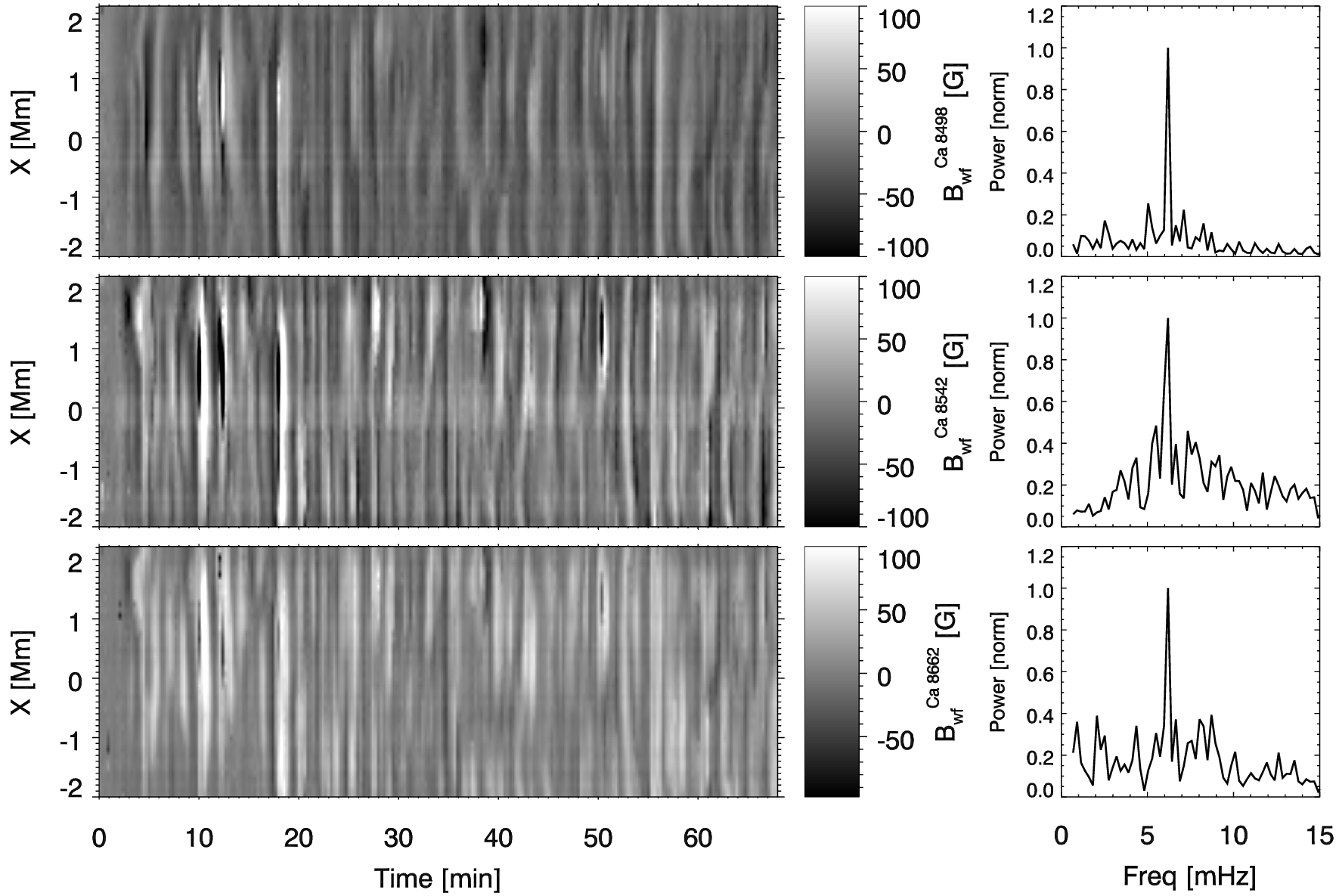}
  \caption{Magnetic field fluctuations spatio-temporal maps obtained using the weak field approximation with the \CaII\ triplet lines (left panels) and power spectra (right panels). Top panels: \CaII\ $\lambda$ 8498.02 \AA\ line,  middle panels: \CaII\ $\lambda$ 8542.09 \AA\ line, bottom panels: \CaII\ $\lambda$ 8662.14 \AA\ line.}
  \label{fig:magnetic_field_Ca}
\end{figure*}

Magnetic field oscillations retrieved from the weak field approximation using \CaII\ IR triplet lines were measured following the same method discussed in Section \ref{sect:magnetic field_Fe} for the \FeI\ $\lambda$ 6301.5 line. In the case of the three \CaII\ lines, we chose the index $i$ in Equation \ref{eq:phi_weak_field} to cover all the wavelengths not farther than 1 \AA\ from the line core at rest. The wavelengths included in the summation are indicated in Figure \ref{fig:Ca_stokes_plot} by the region between the two thick vertical dashed lines at each plot. Figure \ref{fig:magnetic_field_Ca} shows spatio-temporal maps of the fluctuations of the inferred magnetic field oscillations and their power spectra. The fluctuations were obtained as the difference between the inferred magnetic field at each time step and that obtained at the initial time, that is, for the static background without perturbations.

\begin{figure}[!ht] 
 \centering
 \includegraphics[width=9cm]{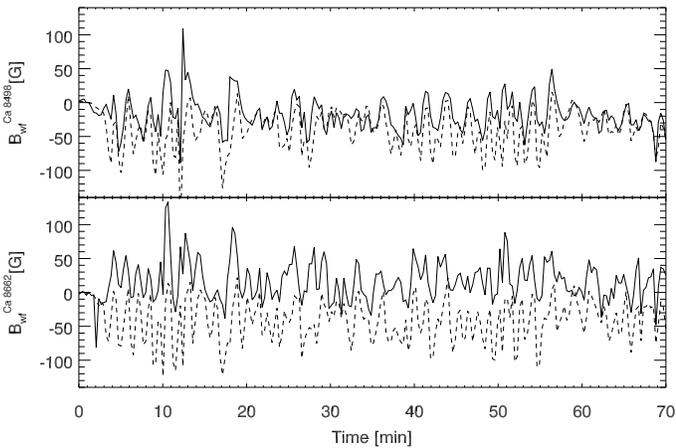}
  \caption{Magnetic field fluctuation obtained using the weak field approximation with the \CaII\ $\lambda$ 8498.02 \AA\ line (top panel) and the \CaII\ $\lambda$ 8662.14 \AA\ line (bottom panels). Solid lines represent the measurement obtained using all the wavelengths closer than 1 \AA\ from the core center (region between thick vertical dashed lines in Figure \ref{fig:Ca_stokes_plot}, while in the dashed line calculation the wavelengths at less than 0.05 \AA\ from the line core were excluded (region between thin vertical dashed lines in Figure \ref{fig:Ca_stokes_plot}).}
  \label{fig:magnetic_field_Ca_plot}
\end{figure}

The temporal evolution of the inferred fluctuations of the magnetic field at a single point for \CaII\ $\lambda$ 8498.02 \AA\ and \CaII\ $\lambda$ 8662.14 \AA\ lines are plotted in Figure \ref{fig:magnetic_field_Ca_plot} (solid lines). They show several surprising features. First, their amplitude is even higher than 100 G. These magnetic field variation is an order of magnitude higher than the highest intrinsic oscillations found in the computational domain. As can be seen in Figures \ref{fig:maps_photosphere} and \ref{fig:maps_chromosphere} and Table \ref{tab:simulations}, the amplitude of photospheric magnetic field oscillations is around 10 G, while at the chromospere it is around 1 G. Secondly, the power of the inferred magnetic field oscillations is concentrated at around 6 mHz, while the power of the real magnetic field oscillations is located in the 5 minutes band at all atmospheric heights (Figures \ref{fig:maps_photosphere} and \ref{fig:maps_chromosphere}). This two facts indicate that the magnetic field retrieved from applying the weak field approximation to the \CaII\ IR triplet does not provide information about the intrinsic oscillations of the magnetic field at any atmospheric layer.    

As a next step, we have performed a new estimation of the magnetic field fluctuations using the weak field approximation, but in this case we have excluded the inner spectral region of the line in the calculations. Wavelengths at less than 0.05 \AA\ from the line core were not included in Equation \ref{eq:phi_weak_field}. Results are plotted with dashed lines in Figure \ref{fig:magnetic_field_Ca_plot}. They show mainly negative fluctuations in the magnetic field with amplitudes around 100 G for the \CaII\ $\lambda$ 8662.14 line and a bit smaller for the \CaII\ $\lambda$ 8498.02 line. Note that the variations are measured with respect to the magnetic field inferred for the background equilibrium atmosphere. Oscillations only produce magnetic field changes between the initial value and around 100 G lower magnetic field. The frequency of these fluctuations indicates that they must be related with velocity, density, or temperature oscillations and, thus, they may be produced by opacity changes which produces variations in the response functions of the \CaII\ lines. Left panels of Figure \ref{fig:shock} show the response function of the Stokes V parameters of the \CaII\ $\lambda$ 8662.14 \AA\ line. Before the shocks the main contribution to the line is at the photosphere, but the dramatic changes produced in the atmosphere by the shock generates a new chromospheric component, whose maximum contribution height travels with the wavefront. The height difference between these two components is around 1 Mm. As the wavefronts travel through the atmosphere, the maximum contribution height to the Stokes V signal of the \CaII\ lines alternates between the photosphere and the chromosphere. Due to the vertical gradient of the magnetic field, the spectral line ``sees'' very different magnetic field strengths. In our umbra model, the variation of the magnetic field between these two layers is around 95 G. It explains the amplitude of the inferred magnetic field fluctuations, and also the fact that oscillations only produce a reduction in the magnetic field with respect to the static state (when the response function peaks at the photosphere and, thus, inferred magnetic field is maximum).      

Finally, we have also computed an estimation of the magnetic field using only the spectral region discarded in the previous calculation (not shown in the figures). The total inferred magnetic field at the initial state is very low, around half the magnetic field strength of the model. When the waves reach high layers the inferred magnetic field shows strong fluctuations, with amplitudes higher than 600 G. These magnitudes neither correspond to any physical property of the sunspot model nor oscillatory behavior, indicating that applying the weak field approximation at the core of the \CaII\ IR lines does not provide useful information about atmospheric configuration and dynamic.

\section{Discussion and conclusions}
\label{sect:conclusions}

In this paper we have presented the synthesis of the Stokes parameters of some spectral lines commonly used in observations from a numerical simulation of wave propagation in a sunspot umbra. This approach allows us to compare directly the variations in the Stokes signal produced by oscillations with the actual atmospheric dynamic, and provide a strong support for the interpretation of spectropolarimetric data in actual observations.   
     
Our analysis include the photospheric \FeI\ $\lambda$ 6301.5 \AA\ line. Phase difference between Doppler velocity and core intensity oscillations ($\Delta\Phi(v_I,I_{\rm core})$) shows 90$^o$ shifts for frequencies below 5 mHz and lower values between -10$^o$ and 30$^o$ for higher frequencies. This measurement is consistent with the phase relation produced if the intensity oscillations are due to opacity fluctuations (in combination with a temperature gradient) rather than intrinsic temperature oscillations, except for frequencies between 5 and 6.5 mHz, where temperature oscillations present higher amplitudes and have an effect on the intensity of the line core. As seen in Figure \ref{fig:dfase_vz_I}, we have obtained this conclusion from two different approaches. First, we have compared the phase difference between Doppler velocity and core intensity with that produced by a model of vertical propagation of slow magnetoacoustic waves. In the later we only included the effect of the opacity fluctuations in the intensity, discarding the temperature variations associated to wave propagation. Phase relation of the theoretical model mimics the measurements from the analysis of the synthetic data, except for the frequency range between 5 and 10 mHz. Second, we have retrieved the temperature at constant optical depth, choosing the optical depth where the \FeI\ $\lambda$ 6301.5 \AA\ line is formed, and we have computed its phase difference with the velocity signal. In the case where we only include the background temperature in the calculation, the phase difference reproduces $\Delta\Phi(v_I,I_{\rm core})$ at all frequencies but the 3 minutes band where intrinsic temperature oscillations are stronger. Finally, including the total temperature (temperature oscillations associated to wave propagation in addition to the background stratification) at constant optical depth provides a quantitative agreement with $\Delta\Phi(v_I,I_{\rm core})$ at all analyzed frequencies. 

Photospheric magnetic field oscillations were inferred by applying the weak field approximation to the \FeI\ $\lambda$ 6301.5 \AA\ line. This method allows us to detect the intrinsic magnetic field fluctuations above noise level (Figures \ref{fig:magnetic_field_fe} and \ref{fig:magnetic_field_fe_noise}), although their amplitude is overestimated. Previous observational works \citep[\eg,][]{BellotRubio+etal2000} found that inferred magnetic field oscillations are caused by opacity fluctuations. This discrepancy is due to the differences in the field strength gradient of the studied sunspots. In those cases where the gradient is strong, opacity fluctuations move upward and downward the formation height of the spectral line, and the intrinsic fluctuations are masked by the strong contrast between background magnetic field at those heights. Our sunspot model has a significantly smaller magnetic field gradient, closer to a large sunspot \citep{Collados+etal1994}, and intrinsic magnetic field oscillations stand out above the variations due to opacity effects.

We have synthesized and analyzed the \CaII\ IR triplet. When shock waves reach chromospheric layers, their Stokes parameters undergo striking modifications which cause highly asymmetrical profiles. This changes are produced by strong velocity gradients and huge differences in the height where these spectral lines are sensitive to magnetic fields. Figure \ref{fig:shock} shows a detailed plot of the evolution of the Stokes profiles during a shock and the atmospheric variations that cause those changes. Magnetic field was also evaluated using the weak field approximation. In the calculations we did not include the wavelengths at the core of the lines, since they proved to give highly unrealistic magnetic field values. The inferred field strength oscillates between its value in the static atmosphere and reductions around 100 G. This variation is due to the vertical gradient of the magnetic field. The response function height of the \CaII\ IR lines at the wavelengths included in this measurement fluctuates between the photosphere (when wave amplitude is low) and the chromosphere (when wavefronts develop into shocks). The amplitude of the inferred magnetic field oscillations is given by the magnetic field difference between those two heights in the background atmosphere. Thus, this kind of measurement could be used in actual observations to obtain an estimation of the variation of the magnetic field with height.       

The study of wave propagation has a strong impact in our understanding of solar atmospheric dynamics and energy balance. They also serve as independent diagnostics of the atmospheric structure. In order to develop more detailed models of the mechanisms that govern the Sun, it is necessary to improve our knowledge of the formation of the spectral lines of interest and the interpretation of spectropolarimetric observations. The combination of numerical simulations and observational techniques is a promising method to advance towards this goal.

\acknowledgements   
Support for TF was provided by the NASA Heliophysics Division
through projects NNH09CE41C, NNH12CF23C, and NNX14AD42G, and by the Solar
Terrestrial program of the National Science Foundation through grant AGS-1127327.
This work is partially supported by the Spanish Ministry of Science
through projects AYA2010-18029 and AYA2011-24808. This work contributes to
the deliverables identified in FP7 European Research Council grant
agreement 277829, ``Magnetic connectivity through the Solar Partially
Ionized Atmosphere''.

\end{document}